%% file: manuscript.tex
\documentclass{article}
\usepackage[latin1]{inputenc}
\usepackage[style=numeric, maxcitenames=10,maxbibnames=100]{biblatex} % for biblatex
\usepackage{verbatim}
\usepackage{xcolor}
\usepackage{graphicx}
\usepackage{caption}
\usepackage{subcaption} 
\usepackage{times}
\usepackage{amsmath}
\usepackage{amssymb}
\usepackage{slashed} 
\usepackage{array}
\usepackage{pstricks}
\usepackage{pst-coil}
\usepackage[english]{babel}
\usepackage{csquotes}  
\usepackage{times}
\usepackage[T1]{fontenc}
\usepackage{url}
\usepackage{cleveref}

\newcommand{\be}{\begin{equation}}
\newcommand{\ee}{\end{equation}}
\newcommand{\bx}{\begin{xalignat}{2}}
\newcommand{\ex}{\end{xalignat}}
\numberwithin{equation}{section}

\title{Supersymmetric Dyons, Superstrings, and Rotating Wormholes } 
\author{
E. Olszewski  \\ 
 {\em Department of Physics and Physical Oceanography} \\ 
 {\em University of North Carolina at Wilmington} \\  
 {\em Wilmington, North Carolina 28403-5606} \\
{\em email: olszewski@uncw.edu}
}
\date{} 

\begin{document}

\maketitle

\begin{abstract}
\input{input_file_ab}

\end{abstract}
\newpage  

\setcounter{page}{1}
\setcounter{section}{0}
\setcounter{equation}{0}
\section{Introduction}
\label{sect1}
\input{input_file_1}
\section{Dimensional Reduction of $D=10, N=1$ Supersymmetry}
\label{sect2_8}

\input{input_file_2}

\section{Dyons with Spin}
\label{sect2_1}
\input{input_file_3}

\setcounter{equation}{0}
\section{Dyons, Type IIB, and Type I $SO(32)$  Superstring Theory}
\label{sect10}
\input{input_file_4}

\section{T-Duality, Gauge/Gravity Duality, and Wormholes}
\label{sect11}

\input{input_file_5}

\setcounter{equation}{0}
\section{Conclusions}
\label{sect7}
\input{input_file_6}

\section*{Data Availability}
\label{sect8}
\input{input_file_7}
\appendix
\section{Appendix}
\label{sect9}
\input{input_file_ap}

\newpage
\printbibliography 
\end{document}

%% file: input_file_ab.tex
We construct supersymmetric dyon solutions based on the 't Hooft/Polyakov monopole.  We show that these solutions satisfy $\kappa$ symmetry constraints and can, therefore be generalized to supersymmetric solutions of type I $SO(32)$ string theory.  After applying a T-duality transformation to these solutions, we obtain two $D3$-branes connected by a wormhole, embedded in an M5 brane.  We analyze  the geometries of each $D3$-brane for two cases, one corresponding to a dyon with vanishing spin, and the other corresponding to a magnetic monopole with non-vanishing spin.  In the case of vanishing spin, the scalar curvature is finite, everywhere, In the case of non-vanishing spin, we find a frame dragging effect due to the spin. We also find that  the scalar curvature diverges along the spin quantization axis, as $1/\rho^2$, $\rho$ being the cylindrical, radial coordinate defined with respect to the spin axis.   These solutions demonstrate the subtle relationship between the  Yang-Mills and gravitational interactions, i.e.\@ gauge/gravity duality.

%% file: input_file_1.tex
In a previous study we have investigated spin 0 dyons within the context  of type  I $SO(32)$ superstring theory in 10 dimensions \cite{olszewskie15}.  Based on the 't~Hooft/Polyakov monopole we have constructed  dyon solutions which are exact solutions of the non-abelian Dirac-Born-Infeld action and a  Wess-Zumino-like action. After applying a T-duality transformation to the solutions we have obtained solutions corresponding to electrically and magnetically charged wormholes\footnote{For additional information about wormholes and their physical properties, please consult the following references   \cite{etde11, Mazharimousavih11, Sakallii15,Jusufik19,ovguna19}.} which connect two  D3-branes.

 In this study we extend our previous work to include solutions with non-vanishing spin.  Specifically, we have applied supersymmetry transformations to the solutions  obtained previously, yielding  spin 1/2 and spin 1 dyons. We then show that the solutions also preserve a combined $\kappa$-symmetry and supersymmetry, so that  they are also solutions of superstring theory.  After applying  a suitable coordinate/gauge transformation, followed by a T-duality transformation, we obtain  rotating wormhole solutions which are both magnetically and electrically charged. 

We now outline the steps in our analysis. In \cref{sect2_8} we review dimensional reduction of $D=10, N=1$ supersymmetry to $D=6, N=2$ and then to $D=4, N=4$ supersymmetry. This reduction is carried out with the purpose of showing, explicitly,  the connection between dyons in four dimensions and dyons derived from superstrings in ten dimensions. 
In \cref{sect2_1},  we use the results of \cref{sect2_8} to re-interpret the  spin 0, dyon solutions in four spacetime dimensions~\cite{harveyj96},  as  a gauge field dimensionally reduced from ten to  six spacetime dimensions. We then apply supersymmetry transformations to the gauge fields, thereby recasting the the supersymmetric dyon solutions in four dimensions as a $D=6, N=2$  supersymmetric gauge theory. As a corollary of our analysis we extend the work  of Kastor and Na~\cite{kastord99}, which applies to supersymmetric magnetic monopoles, to  include supersymmetric dyons. In \cref{sect10} we show that the solutions obtained in \cref{sect2_1} preserve combined $\kappa$-symmetry and supersymmetry and are therefore solutions of  type IIb superstring theory, which  we, then,  recast as solutions of  type  I $SO(32)$ superstring theory, residing on an M5-brane. 
In \cref{sect11}  we  apply a T-duality transformation to the superstring solutions obtained in \cref{sect10}, reducing the theory from  $D=1+4$ to $D=1+3$. The result is two rotating dyons of equal but opposite charge, each residing on a curved $D3$-brane, connected to one another by a wormhole. Finally, we  present numerical and graphical examples, depicting the scalar curvature and frame dragging effect.

Concerning the system of units and sign conventions, we adhere to the same  conventions as in our previous work~\cite{olszewskie15}.  Specifically, in $D$ dimensions the Levi-Civit\`{a} symbol is $\epsilon_{0 1 2 \ldots D}=1$. Greek letters denote space time indices, i.e.\@ 0, 1, 2, 3. Uncapitalized Roman letters denote either the spatial indices  1, 2, 3\footnote{Alternatively, 3-space coordinates are denoted $x, y, z$, where $x \equiv x^1, y \equiv x^2, \text{and } z \equiv x^3$.} or the indices of  the generators of the  gauge group. Capitalized Roman indices denote indices of  ten spacetime dimensions, i.e.\@ 0, 1, 2, ... 9. The signature of the metric, $\eta_{M N}$, is mostly positive. The gamma matricies satisfy the following relations:  $\Gamma^{M \dagger }=\Gamma_M$.  Also, we  employ Lorentz-Heaviside units of electromagnetism  so that $c=\hbar=\epsilon_0=\mu_0=1$. As a consequence, the Dirac quantization condition is $g_e\:g_m = (4 \pi) n_m/2$, $g_e$  ($g_m$) being the electric  (magnetic) charge, and    $n_m$ being an integer.

%% file: input_file_2.tex
In this section we describe the dimensional reduction of $D=10, N=1$ supersymmetric Yang Mills theory, first to the $D=6, N=2$ theory, then to the $D=4, N=4$ theory.  This reduction is performed, specifically, with the purpose of demonstrating how dyons in $D=4$ can be naturally described as evolving from this dimensional reduction process.

We begin with the $D=10$, $N=1$ supersymmetric  Lagrangian density~\cite{polchinskij298},
\be
\mathcal{L} = -\frac{1}{4} F^a_{M N}  F^{a M N } -i\: \frac{1}{2} \bar{\lambda^a} \Gamma^M \mathcal{D}_M \lambda^a\: , \label{eqa_1}
\ee
where 
\be
F^a_{\mu \nu} = \partial_\mu A^a_{\nu} - \partial_\nu A^a_{\mu} - i g_{D9}\: f^{abc}\:  [A^b_{\mu}, A^c_{\nu}] \:. \label{eq8_1b} 
\ee
The  quantity $g_{D9}$ is the Yang-Mills coupling constant in ten dimensions,\footnote{Note that $g^2_{D9}=g^2_{D3}\: (2\pi)^6\alpha'^3$, where $g_{D3}$ is the Yang-Mills coupling constant in four dimensions, and $\alpha'$ is the string coupling constant.  See Appendix B of reference \cite{olszewskie15}. \label{fn_5}} and $f^{abc}$ are the structure constants of the gauge group. Here, the gaugino field, $\lambda$, is the  supersymmetric partner of the gauge field.
The action is invariant under the  supersymmetric transformations,
\begin{subequations}
\label{eqa_2}
\begin{align}
\delta A^a_M = & -i \bar{\zeta} \Gamma_M \lambda^a   \label{eqa_a} \\ 
\delta \lambda^a = & \frac{1}{2} F^a_{MN} \Gamma^{MN} \zeta,  \label{eqa_b} \:. 
\end{align}
\end{subequations}
where $\Gamma^{MN}= \Gamma^M \Gamma^N - \Gamma^N \Gamma^M$.
The gaugino field $\lambda^a$ and supersymmetric parameter $\zeta$ are  32 component Majorana spinors with positive chirality, i.e.\@  $\bar{\lambda}=(\lambda^a)^ T C$, where $C$ is the charge conjugation matrix, and   $ \Gamma^{(10)} \lambda^a= (+1)\: \lambda^a$, where the chirality matrix $\Gamma^{(10)}= i^{-4}\: \epsilon_{01\ldots 9}\: \Gamma^0 \Gamma^1 \ldots \Gamma^9$.\footnote{The chirality matrix in $D$ dimensions is  $\Gamma^{(D)} \equiv K\: \epsilon_{01\ldots(D-1)}\: \Gamma^0 \Gamma^1 \ldots \Gamma^{D-1}$, where $D=2k+2$ and $K=i^{-k}$ for Minkowski signature and $K=i^{-(k+1)}$ for Euclidean signature.\label{fn_1}} 

Using Noether's theorem we obtain the supercurrent by varying the Lagrangian density with respect to the fields $X (X=A^a_M\: \text{or}\: \lambda^a)$ \cite{martins16}
\be
\zeta J^M + \zeta^\dagger J^{\dagger M} \equiv  \sum_X  \frac{\delta \mathcal{L}}{\delta (\partial_M X)} - K^M\: ,\label{eqa_3}
\ee
where $K^M$ is a function whose divergence is the variation of the Lagrangian density under supersymmetry transformations, i.e.\@ $\partial_M K^M = \delta \mathcal{L}$.  The supercharges, $Q_\alpha$,  are obtained from the supercurrents,
\begin{subequations}
\label{eqa_4}
\begin{align}
Q_\alpha = \int d^9x\: J^0_\alpha  \label{eqa_4a} \\ 
Q^\dagger_\alpha = \int d^9x\: J^{\dagger 0}_\alpha  \label{eqa_4b} \:. 
\end{align}
\end{subequations}
The supercharges, which are the generators of supersymmetry transformations,
\be
[\zeta^\dagger Q^\dagger + \zeta Q, X] = \delta X\:,  \label{eqa_5}
\ee
can be obtained from \cref{eqa_3}.  Alternatively, we can  compare \cref{eqa_5}, directly, to \cref{eqa_2}\: and obtain 
\be
Q_\alpha =  -\frac{1}{2} \int dx^9\: F^a_{MN} (\lambda^{a \dagger}\: \Gamma^0 \Gamma^{MN})_\alpha       \: . \label{eqa_6}
\ee
In deriving \cref{eqa_6} we have used the equal-time, canonical anti-commutation and commutation relations,
\begin{subequations}
\label{eqa_7}
\begin{align}
\{ \lambda^a_\alpha(\vec{x}, t), \lambda^{\dagger b}_\beta(\vec{y}, t)     \} & = \delta^{a b}\: \delta_{\alpha \beta}\: \delta^{(9)}(\vec{x} -\vec{y})  \\ 
[A_M^a(\vec{x}, t),E_N^b(\vec{y}, t)] & =-\eta_{M N}\: \delta^{a b}\: \delta^{(9)}(\vec{x} -\vec{y}) \:. 
\end{align}
\end{subequations}
The  field $E^b_N$ is the canonical momentum conjugate to $A^b_N$,   ($E^b_N= F^b_{0 N}$), and $i\bar{\lambda^a_\alpha \Gamma^0}$ is the canonical momentum conjugate to 
$\lambda^a_\alpha $.

We, now, calculate  the anti-commutator $\{Q_\alpha, Q^\dagger_\beta\}$.  This calculation, though similar to  that of Witten~\cite{wittene78}, differs in that his calculation is based on monopole solutions resulting from a Higgs field embedded in $D=4, N=2$ supersymmetry, whereas this calculation is based on the sequential, dimensional reduction from $D=10, N=1$ supersymmetry to $D=6, N=2$ supersymmetry, and finally to $D=4, N=4$ supersymmetry. Our reason for presenting the calculation is to demonstrate the relationship between dyons in $D=4, N=4$ supersymmetry and superstrings in the type I $SO(32)$ theory.  The anti-commutator is evaluated as
\be
\begin{split}
\zeta_2  \zeta^\dagger_1\{ Q,  Q^\dagger  \} & = [\zeta_2 Q, \zeta^\dagger_1 Q^\dagger]  =
  \\
     & = \zeta_2  \zeta^\dagger_1\:  (-\frac{1}{2})^2\: 2^2  \int dx^9\:  F^a_{MN}\: F^a_{KL}\: \Gamma^0\: \Gamma^M \Gamma^N    \Gamma^{ K \dagger} \Gamma^{ L \dagger}\: \Gamma^{ 0 \dagger}\:  . \label{eqa_14a}
\end{split}
\ee
In evaluating \cref{eqa_14a} it is helpful to organize the terms as follows:  in one group all terms where $\{M,N\}$ and $\{K,L\}$ assume  different values, in a second group where both  $\{M,N\}$ are contracted with $\{K,L\}$, resulting in  terms with no $\Gamma$ matrices, and in a third group where one of $\{M,N\}$  is contracted with one of $\{K,L\}$, resulting in terms with two $\Gamma$ matrices. In the first group terms which contain $\Gamma^0$ or $\Gamma^{0 \dagger}$ vanish  because $\Gamma^{0 \dagger}= -\Gamma^0$.  Each of the remaining terms  can be expressed as a divergence.    Such terms  are, typically, assumed to vanish sufficiently fast at the boundary so that these terms make no contribution; however, these terms will become relevant when we consider dyon solutions and their associated central charges (See \cref{eqa_17b}.). The second group evaluates to $P^0$, the energy, i.e.\@
\be
P_0= -\int dx^9\: \Big( F_0^{aM} F^a_{0M}-\frac{1}{4} \eta_{00}F^a_{MN} F^{aMN}\Big)\: . \label{eqa_16}
\ee
In obtaining this result we have used the fact that $\Gamma^{0 \dagger } = -\Gamma^0$ and assumed that all surface integrals vanish.
The third group comprises terms which contain the product $\Gamma^M \Gamma^N$.  If both $M,N \neq 0$, the term vanishes by symmetry arguments and properties of the gamma matricies.  The only terms which are non-vanishing from this group are those that contain $\Gamma^0 \Gamma^N (\text{or}\: \Gamma^{0 \dagger} \Gamma^N), N \neq 0$.  Each of these terms evaluates to
\be
P_N = \int dx^9\: F_0^{aM} F^a_{NM}\: . \label{eqa_17}
\ee 
Thus,
\be
\{ Q,  Q^\dagger  \} =P_0 + \Gamma^0 \Gamma^N P_N \: . \label{eqa_18}
\ee

In preparation for constructing dyon solutions in four dimensions, we constrain the Majorana spinors  $\zeta$ and  $\lambda^a$ in $D=10$, also, to be states of positive chirality in $D=6$, i.e.\@ $\Gamma^{(6)}\: \chi = (+1)\: \chi, (\chi=\zeta, \lambda^a)$.\footnote{See \cref{fn_1} on \Cpageref{fn_1}.}  We, next, re-express the spinors $\chi$ in terms of projections, i.e.\@
\begin{subequations}
\label{eqa_10}
\begin{align}
\chi & = \chi_{+} + \chi_{-}\: , \label{eqa_10a}  \\
\chi_+ &= \chi_{+, +1} + \chi_{+, -1}\: ,  \label{eqa_10b} \\
  \chi_-    & =   \chi_{-, +1} + \chi_{-, -1}\:   . \label{eqa_10c}
\end{align}
\end{subequations}
where
\be
\begin{split}
\chi_{+, \pm 1} = &  \bigg(\frac{1+\Gamma^{(10)}}{2}\: \frac{1+\Gamma^{(6)}}{2} \bigg) \bigg(    \frac{1 + \Gamma'^{(6)}}{2} \bigg)
      \frac{1\pm \Gamma^0 \Gamma^{4}}{2}\: \chi      \\
\chi_{-, \pm 1} = & \bigg(\frac{1+\Gamma^{(10)}}{2}\: \frac{1+\Gamma^{(6)}}{2}\bigg)  \bigg(    \frac{1-\Gamma'^{(6)}}{2} \bigg)
      \frac{1 \pm \Gamma^0 \Gamma^{4}}{2}\: \chi       
\end{split}
\label{eqa_11}
\ee
Here, $\Gamma'^{(6)}$ is the  chirality matrix for dimensions 0, 1, 4, 5, 6, 7,
\be
\Gamma'^{(6)} = \frac{1}{i^2} \epsilon_{014567}\:\Gamma^0 \Gamma^1 \Gamma^4 \Gamma^5 \Gamma^6 \Gamma^7  \:.    \label{eqa_12}
\ee

 We note, in particular, that the $\zeta_{\pm , \pm 1}$, in the s-basis~\cite{polchinskij298}, are 
\be
\begin{split}
\zeta_{+, \pm 1}= &
a^*_{+, \pm 1}  \Bigg[ \begin{pmatrix}
1 \\
0
\end{pmatrix}
\begin{pmatrix}
1 \\
0
\end{pmatrix} 
\begin{pmatrix}
1 \\        
0       
\end{pmatrix}
\begin{pmatrix}
1 \\
0
\end{pmatrix}
\begin{pmatrix}
1 \\
0
\end{pmatrix}  \\
& \pm \begin{pmatrix}
0 \\
1
\end{pmatrix}
 \begin{pmatrix}
1 \\
0
\end{pmatrix} 
\begin{pmatrix}
0 \\    
1         
\end{pmatrix}
\begin{pmatrix}
1 \\
0
\end{pmatrix}
\begin{pmatrix}
1 \\
0
\end{pmatrix} \Bigg]  \\
& + a_{+, \pm 1}  \Bigg[ \begin{pmatrix}
1 \\
0
\end{pmatrix}
\begin{pmatrix}
0 \\
1
\end{pmatrix} 
\begin{pmatrix}
0 \\            
1                
\end{pmatrix}
\begin{pmatrix}
0 \\
1
\end{pmatrix}
\begin{pmatrix}
0 \\
1
\end{pmatrix}   \\
& \pm \begin{pmatrix}
0 \\
1
\end{pmatrix}
\begin{pmatrix}
0 \\
1
\end{pmatrix} 
\begin{pmatrix}
1 \\          
0            
\end{pmatrix}
\begin{pmatrix}
0 \\
1
\end{pmatrix}
\begin{pmatrix}
0 \\
1
\end{pmatrix} \Bigg] \\
\zeta_{-, \pm 1}= &
a^*_{-, \pm 1}  \Bigg[ \begin{pmatrix}
1 \\
0
\end{pmatrix}
\begin{pmatrix}
1 \\
0
\end{pmatrix} 
\begin{pmatrix}
1 \\               
0              
\end{pmatrix}
\begin{pmatrix}
0 \\
1
\end{pmatrix}
\begin{pmatrix}
0 \\
1
\end{pmatrix}  \\
& \pm \begin{pmatrix}
0 \\
1
\end{pmatrix}
 \begin{pmatrix}
1 \\
0
\end{pmatrix} 
\begin{pmatrix}
0 \\           
1              
\end{pmatrix}
\begin{pmatrix}
0 \\
1
\end{pmatrix}
\begin{pmatrix}
0 \\
1
\end{pmatrix} \Bigg]  \\
& + a_{-, \pm 1}  \Bigg[ \begin{pmatrix}
1 \\
0
\end{pmatrix}
\begin{pmatrix}
0 \\
1
\end{pmatrix} 
\begin{pmatrix}
0 \\            
1                
\end{pmatrix}
\begin{pmatrix}
1 \\
0
\end{pmatrix}
\begin{pmatrix}
1 \\
0
\end{pmatrix}   \\
& \pm \begin{pmatrix}
0 \\
1
\end{pmatrix}
\begin{pmatrix}
0 \\
1
\end{pmatrix} 
\begin{pmatrix}
1 \\                
0                   
\end{pmatrix}
\begin{pmatrix}
1 \\
0
\end{pmatrix}
\begin{pmatrix}
1 \\
0
\end{pmatrix} \Bigg]\: ,
\end{split}
\label{eqa_13}
\ee
$a_{+, \pm 1} $ and $a_{-, \pm 1} $ being arbitrary complex constants.

With foresight, we make the following assumptions:
\begin{enumerate}
\item all potential functions $A^a_M=A^a_M(x^i)$, i.e.\@ depend only on the three space coordinates and are time independent, 
\item $A^a_6=A^a_7=A^a_8=A^a_9=0$,  
\item $A_4$ and $A_5$ may or may not commute, 
\item $A_5^a$ asymptotically approaches a non-vanishing vacuum state, while $A_4^a$   may vanish asymptotically, i.e. 
\begin{subequations}
\label{eqa_7a}
\begin{align}
 \lim_{r \rightarrow \infty }A^a_4 A^a_4& =  v^2 \cos^2 \psi \label{eqa_7a_2} \\
 \lim_{r \rightarrow \infty }A^a_5 A^a_5& =  v^2 \sin^2 \psi\:, \label{eqa_7a_3}
\end{align}
\end{subequations}
for $0  < \psi \le \pi/2 $ and $v$ non-vanishing.
\end{enumerate}

The reduction from  ten to six dimensions is trivial.  Since $A^a_6$ through $A^a_9$ vanish, only the gamma matrices $\Gamma^0$ through $\Gamma^5$ appear in the supercharges.  In reducing from ten to six dimensions, the ten dimensional gamma matrices may be represented as a direct product of six dimensional gamma matrices and a four dimensional identity matrix, i.e.\:  $\Gamma^N \times I_4$, where $N=0, \dotsc , 5$.  The gamma matrices act on the first three component spinors of $\chi$, while the four dimensional identity matrix acts on the remaining two.  The only significant consequence of the dimensional reduction is that the spinor $\chi$ is replaced by two spinors
\begin{subequations}
\label{eq2_1}
\begin{align}
 \chi_+ & =  \chi_{+, +1} +  \chi_{+, -1}   \label{eq2_1a} \\
 \chi_- & =  \chi_{-, +1} +  \chi_{-, -1}\: , \label{eq2_1b}
\end{align}
\end{subequations}
and correspondingly the supercharge $Q$ to two supercharges
\begin{subequations}
\label{eq2_2}
\begin{align}
Q_+ & =  Q_{+, +1} +  Q_{+, -1}   \label{eq2_2a} \\
Q_- & =  Q_{-, +1} +  Q_{-, -1}\: . \label{eq2_2b}
\end{align}
\end{subequations}
Thus, dimensional reduction results in a transitioning from $D=10, N=1$ supersymmetry to $D=6, N=2$ supersymmetry with both supercharges being eigenstates of  positive chirality in six dimensions, i.e.\@ $\Gamma^{(6)}\: Q_\pm = + Q_\pm$. 
The central charges are derived from two groups of terms in the anticommutator, the first group and the third group.  A typical non-vanishing boundary term from the first group derives from
\be
F^a_{i j}\: F^a_{kN}\:  \Gamma^i \Gamma^j    \Gamma^{ k \dagger} \Gamma^{ N \dagger}\: , \label{eqa_17a}
\ee
where $N=4,5$.  Boundary terms derived from $F^a_{i j}\: F^a_{4 5}$ involve  a curl integrated over a surface at infinity.  Such terms, which can be expressed as a line integral, vanish asymptotically if $F^a_{4 i}$ and $F^a_{5 i}$ approach zero faster than $1/r$ as $r \rightarrow \infty$.  This is case for monopole or dyon solutions which asymptotically approach zero as $1/r^2$.  The  remaining terms can be expressed as a divergence which becomes a surface integral at the boundary.  If $F^a_{i j}$ approach zero as $1/r^2$ as $r \rightarrow \infty$ as is the case for monopole and dyon solutions, then the surface integral is non-vanishing.  Specifically, the contribution from the first group of terms is 
\be
\Gamma^1 \Gamma^2    \Gamma^{ 3 }  \Gamma^{ 4 }\: g_m v \: \cos \psi   + \Gamma^1 \Gamma^2    \Gamma^{ 3 }  \Gamma^{ 5 }\: g_m v \sin \psi\:  , \label{eqa_17b}
\ee
where the magnetic charge $g_m$ is obtained from the relationship
\be
g_m v\sin \psi = \int_{S_\infty}  B^{a i} A^a_5\: dS_i\: .  \label{eqa_17c}
\ee
We have used the fact that the asymptotic behavior of $A^a_4$ is given by \cref{eqa_7a_2}, and that the magnetic field is given by
\be
B^{a k} = \frac{\epsilon^{kij}}{2!} F^a_{i j}\: . \label{eqa_18b}
\ee 
In obtaining \cref{eqa_17c} we have used 
\be
 \Gamma^i \Gamma^j    \Gamma^k \Gamma^{ 4} \partial_k (F^a_{ij }\: A^a_{4}) = \Gamma^i \Gamma^j    \Gamma^k \Gamma^{ 5} (F^a_{i j}\: F^a_{k 4} +   \frac{\epsilon^{kij}}{2!} \mathcal{D}_k F^a_{ij }  A^{a}_4)\: . \label{eqa_18c}
\ee
In \cref{eqa_18c} the second term to the right of the equal sign vanishes by virtue of the equations of motion, specifically that the divergence of the magnetic field vanishes.

The contribution to the central charges from the third group of term corresponds to the momentum in the $x^4$ and $x^5$ directions. The relevant terms from \cref{eqa_14a}  
\be
\int dx^6\:\int dx^3\: F^a_{0K}\: F^a_{KN}\: \Gamma^0 \Gamma^0\:  \Gamma^N\:\Gamma^{0 \dagger} , \label{eqa_19}
\ee
where $N=4, 5$. The portion of the integral over the six dimensional space yields the volume of the six dimensional space which we normalize to one.  The remaining part of the integral can be expressed as a divergence which by virtue of \cref{eqa_7a_3} yields a non-vanishing surface contribution.  Substituting  the following expression 
\be
\Gamma^0  \Gamma^N\: \partial_k (F^a_{0K}\: A^a_{N}) =\Gamma^0  \Gamma^N\: ( F^a_{0K}\: F^a_{KN}  + \mathcal{D}_k F^{aK}_{0}\: A^a_{N})\: \label{eqa_20} 
\ee
into \cref{eqa_19} and using the fact the last term in \cref{eqa_20} vanishes by virtue of the equations of motion, i.e.\@  the divergence of the electric field vanishes, we obtain the additional contributions to the central charges,
\be
\Gamma^0 \Gamma^4\: g_e v \: \cos \psi   + \Gamma^0 \Gamma^{ 5 }\: g_e v \sin \psi\:  , \label{eqa_21}
\ee
 Here, we have used that the electric charge is obtained
\be
g_e v \cos \psi =\int_{S_\infty} E^{ai} A^a_4\: dS_i\: , \label{eqa_22}
\ee
where $E^{ai}=E^{a}_i= F^{a}_{i0}$.  Substituting \cref{eqa_17b} and \cref{eqa_22} into \cref{eqa_18}, we obtain

\be
\begin{split}
\{ Q_\mathfrak{a},  Q_\mathfrak{b}^\dagger  \} = & \delta_{\mathfrak{a} \mathfrak{b}} \{P_0 + \Gamma^0 \Gamma^i P_i \\
              & + \Gamma^0 \Gamma^4\: g_e v \: \cos \psi   + \Gamma^0 \Gamma^{ 5 }\: g_e v \sin \psi \\
							& + \Gamma^1 \Gamma^2    \Gamma^{ 3 }  \Gamma^{ 4 }\: g_m v \: \cos \psi   + \Gamma^1 \Gamma^2    \Gamma^{ 3 }  \Gamma^{ 5 }\: g_m v \sin \psi \}\: , \label{eqa_23}
\end{split}
\ee
for $(\mathfrak{a}, \mathfrak{b} = +, -)$.\footnote{We use fraktur font to denote '+' or '-.'\label{fn_2}}  Simplifing the terms involving central charges we obtain
\be
\begin{split}
\{ Q_\mathfrak{a},  Q_\mathfrak{b}^\dagger  \} = & \delta_{\mathfrak{a} \mathfrak{b}} \{P_0 + \Gamma^0 \Gamma^i P_i 
               + g\: v\:  \exp(i\Gamma^{(4)} \psi'\: \Gamma^0 \Gamma^{ 4 }) \\ 
								& (\cos \psi +i\Gamma^{(4)} \sin \psi\: \Gamma^0 \Gamma^{ 4 }\:  \Gamma^{ (6) })\: \Gamma^{(6)} \Gamma^0 \Gamma^{ 4 }\} \: , \label{eqa_24}
\end{split}
\ee
The charge $g$ and the angle $\psi'$ are defined by\footnote{Because of our choice of metric, i.e.\@  $\eta_{00}=-1$, electromagnetic duality implies $ ^*B^a \rightarrow E^a$  and $ ^*E^a \rightarrow -B^a$.  \label{ftn_3c} }
\be
\begin{aligned}
g &= \sqrt{g_m^2+g_e^2} \\
\tan \psi' & = \frac{g_m}{g_e}\: .  \label{eqa_25}
\end{aligned}
\ee
Since   $\Gamma^{(6)} \chi = 1\:\: \chi$, we can simplify  \cref{eqa_24}
 \be
\{ Q_\mathfrak{a},  Q_\mathfrak{b}^\dagger  \} =  \delta_{\mathfrak{a} \mathfrak{b} } (P_0 + \Gamma^0 \Gamma^i P_i +  g\: v\:( \exp\{i\Gamma^{(4)} (\psi'+\psi )\Gamma^0 \Gamma^4\}\Gamma^0 \Gamma^4. \label{eqa_26}
\ee

Before reducing from six to four dimensions, we note that
\be
Q_{\mathfrak{a}, \pm 1} = \frac{(1 \pm \Gamma^0 \Gamma^{ 4 } )}{2} Q\mathfrak{a}\: . \label{eqa_27}
\ee
In the rest frame of the system, i.e.\@ $P_i=0$ and $P_0=M$, $M$ being the rest energy of the system, we can show
\be
\begin{aligned}
\{ Q_{\mathfrak{a}, +1},  Q_{\mathfrak{b}, +1}^\dagger  \} & =  \delta_{\mathfrak{a} \mathfrak{b} } (M   + g\: v \exp\{i\Gamma^{(4)} (\psi'+\psi)\}) \\
\{ Q_{\mathfrak{a}, -1},  Q_{\mathfrak{b}, -1}^\dagger  \} & =  \delta_{\mathfrak{a} \mathfrak{b} }(M   -  g\: v \exp\{-i\Gamma^{(4)} (\psi'+\psi)\})\:  \\
\{ Q_{\mathfrak{a}, +1},  Q_{\mathfrak{b}, -1}^\dagger  \} & = 0
\: . \label{eqa_28} 
\end{aligned}
\ee
Alternatively, we define
\be
\begin{aligned}
 Q_\mathfrak{a}^1 &=\frac{(1 +  \Gamma^{ 4 } )}{2} Q_{\mathfrak{a}, +1} \\
Q_\mathfrak{a}^2 &=\frac{(1 - \Gamma^{ 4} )}{2} Q_{\mathfrak{a}, -1}
\: . \label{eqa_29}
\end{aligned}
\ee
We can show by direct substitution of \cref{eqa_29} into \cref{eqa_28} that
\be
\{ Q_\mathfrak{a}^i,   Q_\mathfrak{b}^{j  \dagger}  \}  =\delta_{\mathfrak{a} \mathfrak{b}} (\delta_{i j}\: M +  \Gamma^4\:  g\: v \exp\{-i\Gamma^{(4)} (\psi'+\psi)\})\: , \label{eqa_30}
\ee
for $i, j = 1, 2$.
The reduction from six to four dimensions is relatively straightforward.  In reducing from ten to six to four dimensions the requisite ten dimensional gamma matrices are represented
\be
\begin{aligned}
\Gamma^\mu &=\gamma^\mu \times I_2 \times I_4 \\
\Gamma^{(4)}  &=\gamma^5 \times I_2 \times I_4 \\
\Gamma^4  &=\gamma^5 \times \sigma_1 \times I_4   \\
\Gamma^5  &=\gamma^5 \times \sigma_2 \times I_4\:
\label{eqa_31} 
\end{aligned}
\ee
Here, $\gamma^\mu$ and $\gamma^5$ are the four dimensional gamma matrices, $\sigma_1$ and $\sigma_2$ are Pauli matrices, and $I_2$ and $I_4$ are the identity matrices in two and four dimensions, respectively.  Finally, the reduction from six dimensions to four dimensions requires that $\Gamma^4$ and $\Gamma^{(4)}$ from \cref{eqa_31} be substituted into \cref{eqa_30}.  In reducing from $D=6, N=2$ to $D=4, N=4$ supersymmetry, each supercharge $Q_\mathfrak{a}$ is replaced by two supercharges $Q_{\mathfrak{a}, \pm 1}$.  

The supersymmetry algebra, \cref{eqa_30}, obtained from dimensional reduction of $D=10,N=1$ supersymmetry, differs from that of Witten and Olive~\cite{wittene78} which is based on  $D=4,N=2$ supersymmetry.   The most obvious distinction is that there are two sets of supercharges, i.e.\@ $ (\mathfrak{a} = +, -)$. In our construction of dyons with spin in \cref{sect2_1}, the second set of  supercharges  generates  spin 1, dyon solutions in addition to  spin 1/2 and spin 0 solutions. The other distinction derives from the fact  that  the components of the vector potential,  $A^a_4, A^a_5$ in our analysis,  replace the components of the Higgs field, in Witten's analysis. Witten removes one of these components of the Higgs field by performing a chiral rotation, which would, in a certain sense, be equivalent to setting $\psi = 0$, in our analysis. 
In our subsequent analysis of dyons with spin, \cref{sect2_1}, we do not eliminate one of the $A^a_4, A^a_5$ by a coordinate rotation, analogous to the chiral rotation. The reason is that our analysis is complicated because the $A^a_4$ and $A^a_5$, in general, do not commute.    Instead, we are able to set $\psi^\prime = \psi$, which is a direct consequence of the dyon solutions being BPS states.

%% file: input_file_3.tex
In this section we review the construction of dyons with spin.  One method of incorporating spin is to construct dyon solutions from the $D=6$, supersymmetric extension of the Yang-Mills-Higgs action.  This methodology  shows, implicitly, the relationship between dyons with spin in $D=4$ and superstrings.  
We begin the analysis with a discussion of 't Hooft/Polyakov monopole which is derived from the the Yang-Mills-Higgs Lagrangian density.  't Hooft~\cite{thooftg76} and Polyakov\cite{polyakova74} have shown that  within the context of the spontaneously broken, Yang-Mills gauge theory SO(3) magnetic monopole solutions of finite mass must necessarily exist and furthermore possess an internal structure. These solutions, which possess zero spin, are derived from the  Yang-Mills-Higgs Lagrangian,
\be
\mathcal{L} = -\frac{1}{4} F^a_{\mu \nu}  F^{ \mu \nu a} + \frac{1}{2} \mathcal{D}_\mu \Phi^a  \mathcal{D}^\mu \Phi^a - V(\Phi^a \Phi^a)      \: ,\label{eq22_1}
\ee
where 
\be
F^a_{\mu \nu} = \partial_\mu A^a_{\nu} - \partial_\nu A^a_{\mu} - i g_{D3}\: f^{abc}\:  [A^b_{\mu}, A^c_{\nu}] \:. \label{eq22_1b} 
\ee
The Higgs field $\Phi^a$ is a scalar transforming according to the adjoint representation of the gauge group, and consequently, its covariant derivative is
\be
\mathcal{D}_\mu \Phi^a = \partial_\mu \Phi^a  - i g_{D3}\: f^{abc}[A^b_{\mu}, \Phi^c]  \:.  \label{eq22_1c}
\ee
The  quantity $g_{D3}$ is the Yang-Mills coupling constant in four dimensions, and $f^{abc}$ are the structure constants of the gauge group.  For our purposes we assume that the gauge group is $SU(2)$ (or a group which contains $SU(2)$ as a subgroup).  In addition, we require that  the potential $V(\Phi^a \Phi^a)$ vanishes so that  the magnetic monopole solutions are BPS states,  which are solvable in closed form~\cite{harveyj96,kastord99,olszewskie12,olszewskie15}.  Straightforwardly, one can also show that these solutions can be, modified to be electrically charged, as well as magnetically charged. As a consequence of the solutions being BPS states, one can show that the electric and magnetic component of the fields are related to $\Phi^a$,
\begin{subequations}
\label{eq22_3}
\begin{align}
E^a_i  & =   \cos \psi\: \mathcal{D}_i  \Phi^a \\
B^a_i & =     \sin \psi\: \mathcal{D}_i  \Phi^a \:, 
\end{align}
\end{subequations}
where  
\begin{subequations}
\label{eq22_4}
\begin{align}
E^a_i= & F^a_{i 0} \\
B^a_i = &  \epsilon_i^{jk}F^a_{jk} \:. 
\end{align}
\end{subequations}
The electric and magnetic fields are obtained from $E^a_i$ and $B^a_i$, 
\begin{subequations}
\label{eq22_5}
\begin{align}
E_i= & E^a_i \frac{\Phi^a}{v} \\
B_i= & B^a_i \frac{\Phi^a}{ v}  \:. 
\end{align}
\end{subequations}
Here,
\be
v^2 = \lim_{r \rightarrow \infty} \Phi^a \Phi^a\: . \label{eq3_4}
\ee
See \cref{eq3_2} below.

In \cref{eq22_3} the electric, $q_e$, and magnetic, $q_m$, charges are
\begin{subequations}
\label{eq22_6}
\begin{align}
q_e = & q \cos \psi  \\
q_m = & q \sin \psi   \:, 
\end{align}
\end{subequations}
where $q = \sqrt{q_e^2+q_m^2}$. For these solutions $\psi'=\psi$ (See \cref{eqa_25}.).

From the perspective of six dimensions the function $\Phi^a$ can be reinterpreted as gauge fields
\begin{subequations}
\label{eq3_1}
\begin{align}
A^a_4 & =  \Phi^a \cos \psi  \label{eq3_1a} \\
A^a_5 & =  \Phi^a \sin \psi \label{eq3_1b} \:. 
\end{align}
\end{subequations}
This follows because the Higgs field $\Phi^a$  does not depend on the coordinates of dimensions four and five so that 
 under gauge transformations,  the components $A^a_4$ and $A^a_5$ transform in the same manner as $\Phi^a$.  
In six dimensions the dyon is described in terms of  the potential function 
\begin{equation}
\begin{split}
A  = & A_\mu dx^\mu + A_4 dx^4 + A_5 dx^5 \\
  = &   \cos \psi\: v\:   Q(r)\:   T^{r}\: dt  + \frac{W(r)}{g_{D3}}\: [T^{\theta} \sin \theta \:n_m\: d\phi - T^{\phi}\: d\theta]  \\     
                                 & +  \cos \psi\: v\:  Q(r)\:   T^{r}\: dx^4    +  \sin \psi\: v\:   Q(r)\:  T^{r}\: dx^5 \:,  \:\label{eq2_23}
\end{split}
\end{equation}
where $v$ is  vacuum expectation value of $\Phi^a$ in the asymptotic limit of large $r$ (See \cref{eqa_7a_2}.). The magnetic charge of the dyon is $g_m=4\pi\: n_m/g_{D3}$, for $n_m$ an integer, which is the Higgs field winding number.   
The $T^r, T^\theta, T^\phi$,  constitute a representation of the $SU(2)$ algebra. The quantities $r, \theta, \phi$ are the spherical polar coordinates in three dimensions.~\footnote{In the transformation to spherical polar coordinates, we have chosen the $x$-axis, rather that the $z$-axis, to be  the azimuthal axis.  The motivation  for this choice is to provide consistency with our choice of $\Gamma$ matrices.  Specifically, spin states are chosen to be eigenvalues of the spin operator $S^x$.  See \cref{eq3_39}. } The elements $T^{r}, T^{\theta}, T^{\phi}$ are related to $T^a, (a= 1, 2, 3)$,
\begin{subequations}
\label{eq2_25c}
\begin{align}
T^{r} \equiv \mathbf{T} \cdot \mathbf{e}_r= T^a e_r^a = & \:T^y \:\sin \theta \cos n_m \phi + T^z  \:\sin \theta \sin n_m \phi 
              + T^x \:\cos \theta  \\
T^{\theta} \equiv \mathbf{T} \cdot \mathbf{e}_\theta= T^a e_\theta^a = & \:T^y \:\cos \theta \cos n_m \phi + T^z\:\cos \theta\: \sin n_m\phi - T^x \:\sin \theta \\
 T^{\phi} \equiv \mathbf{T} \cdot \mathbf{e}_\phi= T^a e_\phi^a  = & \:-T^y \:\sin n_m \phi + T^z \:\cos n_m\phi  \:,   
\end{align}
\end{subequations}
where the $T^a$ are generators of an $SU(2)$ subalgebra of $SO(32)$.\footnote{The gauge group $SO(32)$ is relevant for our discussion of superstrings in \cref{sect10}.} 
\begin{subequations}
\label{eq2_125}
\begin{align}
 \mathbf{e}_r=  e_r^a\: \hat{e}_{x^a}= &  \cos \theta\: \hat{e}_{x^1}+ \sin \theta\: \cos n_m \phi\: \hat{e}_{x^2}+ \sin \theta \sin n_m \phi\: \hat{e}_{x^3}
                \\
 \mathbf{e}_\theta=  e_\theta^a\: \hat{e}_{x^a} = &  -\sin \theta\: \hat{e}_{x^1} + \cos \theta\: \cos n_m \phi\: \hat{e}_{x^2} + \cos \theta\: \sin n_m\phi\: \hat{e}_{x^3} \\
 \mathbf{e}_\phi=  e_\phi^a\: \hat{e}_{x^a}  = & \:-\sin n_m \phi\: \hat{e}_{x^2} + \cos n_m\phi\: \hat{e}_{x^3}  \:.   
\end{align}
\end{subequations}
Here the $\hat{e}_{x^a}, (a=1, 2, 3)$ are unit vectors in the $x, y, z$ directions, respectively.

The Higgs field is 
\be
\Phi^a T^a= v\: Q(r) e_r^a\: T^a = v\: Q(r) T^r\: . \label{eq3_2}
\ee
Using \cref{eq2_25c} we can express the $\mathcal{D}(\Phi^a T^a)$ in spherical polar coordinates
\be
\begin{split}
\mathcal{D}_r (\Phi^aT^a) &=   v Q(r)^\prime T^r \\
\mathcal{D}_\theta (\Phi^aT^a) &=  v [1-W(r)]Q(r) T^\theta   \\
\mathcal{D}_\phi (\Phi^aT^a) &=  v [1-W(r)]Q(r)\:  n_m \sin \theta\: T^\phi  
\end{split}
\label{eq2_23b} 
\ee

The solutions $W(r)$ and $Q(r)$ are obtained as in reference~\cite{olszewskie12}
\begin{subequations}
\label{eq2_26}
\begin{align}
W(r) & = w(u)  = 1 - \frac{u}{\sinh u}  \\
 Q(r) & = q(u)  = \coth u - \frac{1}{u}  \:,
\end{align}
\end{subequations}
where 
the dimensionless variable $u$  is related to the radial coordinate $r$,
\be
u= \frac{r}{L_\text{dyon}}  \:. \label{eq2_25}
\ee
The quantity $L_\text{dyon}$ characterizes the size of the dyon, i.e.\@ the region of space in which it exhibits internal structure:
\be
L_\text{dyon} = \frac{1}{ \sin \psi  M_\text{gluon}}\:, \label{eq3_26} 
\ee
where the mass of the gluon, resulting from spontaneous symmetry breaking, is 
\be
M_\text{gluon}= g_{D3}\: v\: . \label{eq3_27}
\ee
In addition, the mass of the dyon is related to the mass of a gluon 
\be
M_\text{dyon} = g v = \frac{g}{g_{D3} } M_\text{gluon}\: . \label{eq3_27b}
\ee

For our purposes we also require that solutions be invariant under $SL(2,Z)$ transformations, weak/strong duality, so that we include in the Lagrangian density  Witten's $\theta$ term\cite{wittene79}
\be
\mathcal{L}_\theta = -\frac{\theta\: g_{D3}^2}{32\pi^2} F^{a}_{\mu\nu} \hspace{.1cm}^*F^{a\mu\nu}.\: \label{eq22_7}
\ee
This term contributes only a surface term to the action, and therefore does not affect the classical equations of motion.  In the monopole sector of the theory, however, the term does have a non-trivial effect in that it shifts the allowed values of the electric charge~\cite{harveyj96}.  The electric charge, $q_e$ is given as
\be
q_e=n_e\: g_{D3} -n_m \frac{g_{D3}\: \theta}{2 \pi}\: , \label{eq3_28}
\ee
$n_e$ being an integer.

The dyon solutions, \cref{eq2_23}, also satisfy the equations of motion derived from the supersymmetric Lagrangian density, \cref{eqa_1} with the gaugino field set equal to zero.  The solutions, \cref{eq3_1} and \cref{eq2_23}, satisfy the assumptions placed on the $D=10$ supersymmetric solutions discussed in \cref{sect2_8}, with the additional property that the solutions are also BPS states.

In order to construct dyon solutions with spin we begin with the $D=6, N=2$ supersymmetric Yang-Mills theory, obtained from the dimensional reduction of the $D=10, N=1$ theory, presented in \cref{sect2_8}. The $D=6, N=2$  theory comprises  two supercharges of positive chirality in six dimensions, $Q_{\mathfrak{a}}, (\mathfrak{a}= +, -)$.  The theory is invariant under supersymmetry transformations generated by supercharges $Q_{\mathfrak{a}}$\footnote{The  gamma matrices in $D=10$ are represented as $\Gamma^N \times I_4$, where $\Gamma^N$ are six dimensional gamma matrices.  See \cref{sect2_8}. \label{fn_4}}
\begin{subequations}
\label{eq3_29}
\begin{align}
\delta A^a_0 = \sum_{\mathfrak{a}}    \delta A^a_{\mathfrak{a} 0} = & -i \bar{\zeta_\mathfrak{a}} \Gamma_0\:  \lambda^a_\mathfrak{a}  \notag \\
	 =  & -i \zeta_{\mathfrak{a}}^\dagger  \lambda^a_{\mathfrak{a}}                 \label{eq3_29a} \\
	\delta A^a_i = \sum_{\mathfrak{a}} \delta A^a_{\mathfrak{a} i} = & -i	  \bar{\zeta_{\mathfrak{a}}}   \Gamma^i   \lambda^a_\mathfrak{a}                   \notag  \\
	 =   & -i \zeta_{\mathfrak{a}}^\dagger \Gamma^0 \Gamma_i\:  \lambda^a_{\mathfrak{a}}  \label{eq3_29b} \\	
\delta A^a_4 = \sum_{\mathfrak{a}}	\delta A^a_{\mathfrak{a} 4} = & -i \bar{\zeta_\mathfrak{a}} \Gamma_4  \lambda^a_\mathfrak{a} \notag  \\
	 =  &       -i \zeta_{\mathfrak{a}}^\dagger \Gamma^0 \Gamma^4 \:  \lambda^a_\mathfrak{a}             \label{eq3_29d} \\
	\delta A^a_5 = \sum_{\mathfrak{a}} \delta A^a_{\mathfrak{a} 5} = & -i \bar{\zeta_\mathfrak{a}} \Gamma_5  \lambda^a_\mathfrak{a} \notag  \\
	= &  - \zeta_{\mathfrak{a}}^\dagger  \Gamma^{(6)}\:  \Gamma^0 \Gamma^4\: \Gamma^{(4)}  \lambda^a_{\mathfrak{a}}                \label{eq3_29c} \\
\delta \lambda^a =   \sum_{\mathfrak{a}}     \delta \lambda^a_\mathfrak{a} = & \sum_{\mathfrak{a}}    \frac{1}{2} F^a_{MN} \Gamma^{MN} \zeta_\mathfrak{a} \notag \\
    =  &  \sum_{\mathfrak{a}}   \bigg( \slashed{E^a} \Gamma^0 -\Gamma^i F^a_{i5} \Gamma^0 (\Gamma^0 \Gamma^5)  \notag \\
		& - i \slashed{B^a} \Gamma^0 \Gamma^{(4)} -i \Gamma^i F^a_{i4} \Gamma^0 \Gamma^{(4)} \Gamma^{(6)}\:  (\Gamma^0 \Gamma^5) \bigg) \zeta_\mathfrak{a} .  \label{eq3_29e}  
\end{align}
\end{subequations}

Supersymmetry is broken by a part of $\zeta_\mathfrak{a}$ which is an eigenstate of $\Gamma^0 \Gamma^4$ with eigenvalue -1, i.e.\@  $\zeta_{\mathfrak{a}, -1}$.
Substituting \cref{eq22_3} and \cref{eq22_4} into     \cref{eq3_29a} and  \cref{eq3_29b}, we obtain
\begin{subequations}
\label{eq3_30}
\begin{align}
\delta A^a_{\mathfrak{a} M} &= 0\: . \label{eq3_30a}  \\
\delta \lambda^a_{\mathfrak{a}, -1}  & =  2 (\slashed{E^a} \Gamma^0 - i \slashed{B^a} \Gamma^0 \Gamma^{(4)})\: \zeta_{\mathfrak{a}, -1}\: , \label{eq3_30b} \\
\delta \lambda^a_{\mathfrak{a}, +1} & =    0\: , \label{eq3_30c} 
\end{align} 
\end{subequations}
As is  is a characteristic of BPS states,  half of the supersymmetries are broken, i.e.\@ for 
$\zeta_{\mathfrak{a}, -1}$, and half are unbroken, i.e.\@ for $\zeta_{\mathfrak{a}, +1}$.  The dimensional reduction to $D=4$ is trivial.  The six dimensional gamma matrices are replaced by those given in \cref{eqa_31}.  It is notable that in our analysis, there are two broken supercharges, a result which differs from those of others. See Harvey, for example, \cite{harveyj96}.  The difference is a consequence of the fact these other analyses begin with $D=4,N=2$  supersymmetric Yang-Mills-Higgs theory.  In contrast, we begin with $D=10, N=1$ supersymmetric Yang-Mills theory with only gauge fields, and through dimensional reduction, obtain a second supercharge.  For these dyon solutions the gaugino field has been explicitly set to zero.  The broken supersymmetry transformations which are generated by the two supercharges, each result in a non-vanishing contribution to the fermion (gaugino) field.  Furthermore,  these transformations which break supersymmetry do not change the energy of the system, so that these non-vanishing fermionic ``zero'' modes can be considered as deformations of the dyon background which keep the energy of the dyon fixed~\cite{harveyj96}.  Since each of these fermionic modes carries spin 1/2, it is possible to construct dyon states, i.e.\@ deformed dyon backgrounds, with either spin 1/2 or spin 1. 

To first order the supersymmetry transformation, \cref{eq3_29a} leaves the potential function , $A^a_M$, unchanged. In reference~\cite{kastord99}, Kastor and Na have shown, because of the non-linearity inherent in the supersymmetry transformations, that there are non-vanishing contributions to $A^a_M$ when higher order corrections to the supersymmetry transformations are taken into account. Their methodology utilizes an interative procedure to calculate higher order corrections to the supersymmetry transformations.  They perform their analysis  using magnetic monopole solutions, i.e.\@ dyons with vanishing electric charge or $\psi=\pi/2$.  Since the changes resulting from the inclusion of electric charge are not immediately obvious, we review their methodology when electric charge is included in the analysis.

They begin with an iterative expansion of the the supersymmetry transformations
\be
\Psi = \exp(\delta)\: \bar{\Psi}= \bar{\Psi} +\delta\bar{\Psi} + \frac{1}{2!} \delta^2 \bar{\Psi}+ \frac{1}{3!} \delta^3 \bar{\Psi}  + \frac{1}{4!} \delta^4 \bar{\Psi} \: , \label{eq3_31}
\ee
where $\Psi$ represents both bosonic and fermionic fields after the transformation, and $\bar{\Psi}$ the bosonic fields before the transformation. This series can be interpreted as follows.  The second term to the right of the second equal sign is obtained directly from \cref{eq3_30}.  The third term is obtained by substituting the second term into \cref{eq3_30}.  The series  terminates after the fourth term because of the Grassman nature of $ \zeta_{\mathfrak{a}, -1}$.  Substituting \cref{eq3_30a} and \cref{eq3_30b} in \cref{eq3_31}, we obtain
\begin{subequations}
\label{eq3_32}
\begin{align}
\delta^2 A^a_{ 0} =  \sum_{\mathfrak{a}} \delta^2 A^a_{\mathfrak{a} 0} &= 2\: \zeta^\dagger_{\mathfrak{a}, -1} \Gamma^0 \Gamma^{(4)} \Gamma^j \zeta_{\mathfrak{a}, -1} B^a_j    \label{eq3_32a} \\
\delta^2 A^a_{ i} =  \sum_{\mathfrak{a}} \delta^2 A^a_{\mathfrak{a} i} &=  -i 2\: \zeta^\dagger_{\mathfrak{a}, -1} \Gamma_i  \Gamma^j \zeta_{\mathfrak{a}, -1} E^a_j   \label{eq3_32b} \\ 
\delta^2 A^a_{ 4} =  \sum_{\mathfrak{a}} \delta^2 A^a_{\mathfrak{a} 4} &=  2\: \zeta^\dagger_{\mathfrak{a}, -1} \Gamma^0 \Gamma^{(4)} \Gamma^j \zeta_{\mathfrak{a}, -1} B^a_j  \label{eq3_32c} \\
\delta^2 A^a_{ 5} =  \sum_{\mathfrak{a}} \delta^2 A^a_{\mathfrak{a} 5} &=    2\: \zeta^\dagger_{\mathfrak{a}, -1} \Gamma^0 \Gamma^{(4)} \Gamma^j \zeta_{\mathfrak{a}, -1} E^a_j  \label{eq3_32d} \\
\delta \lambda^a = \sum_{\mathfrak{a}} \delta \lambda^a_{\mathfrak{a}, -1} & = \sum_{\mathfrak{a}}  2\: (\slashed{E^a} \Gamma^0 - i \slashed{B^a} \Gamma^0 \Gamma^{(4)})\: \zeta_{\mathfrak{a}, -1}\: .  \label{eq3_32e}  
\end{align}
\end{subequations}
Following Kastor and Na~\cite{kastord99}, we evaluate the matrix elements in  \cref{eq3_32}.  We, first, quantize the fermionic zero modes.  This involves replacing the complex constants, $a^*_{-, \pm 1}  $ in $ \zeta_{-, \pm 1}$, \cref{eqa_13}, by the operators $a_{\mathfrak{a}, - 1} $ and $a^\dagger_{\mathfrak{a}, - 1}$, and then integrating the anticommutator of the fermionic zero modes, \cref{eq3_32e}, 
\be
\delta_{ab}\int dx^3 dy^3\: \{ \delta \lambda^a_{\mathfrak{a}, -1}, \delta \lambda^{b \dagger}_{\mathfrak{b}, -1}   \}\: .  \label{eq3_34}
\ee
 Using \cref{eqa_7} and \cref{eqa_13},we obtain 
\begin{subequations}
\label{eq3_33}
\begin{align}
\{  a_{\mathfrak{a},  -1} , a^\dagger_{\mathfrak{b}, -1} \}& = \frac{1}{4 M} \delta_{\mathfrak{a} \mathfrak{b}}\; , \label{eq3_33a} \\
\{  a^\dagger_{\mathfrak{a},  -1} , a^\dagger_{\mathfrak{b}, -1} \} & = 0\: , \label{eq3_33b} \\
\{  a_{\mathfrak{a},  -1} , a_{\mathfrak{b}, -1} \} & = 0\: , \label{eq3_33c}
\end{align}
\end{subequations}
 where we have used the fact that the mass of the dyon is
\be
M=\int dx^3\: (\vec{E}^a \cdot \vec{E}^a + \vec{B}^a \cdot \vec{B}^a) =gv\: . \label{eq3_35}
\ee
Applying \cref{eq3_33} in the evaluation of \cref{eq3_32}, we obtain
\begin{subequations}
\label{eq3_36}
\begin{align}
   \delta^2 A^a_{ 0} &= -2\vec{\mu}_m \cdot \frac{1}{g} \overrightarrow{\mathcal{D}\Phi}^a    \label{eq3_36a} \\
\delta^2 \overrightarrow{A}^a  &=   2\vec{\mu}_e \times \frac{1}{g} \overrightarrow{\mathcal{D}\Phi}^a    \label{eq3_36b} \\ 
\delta^2 A^a_{ 4} &=  2\vec{\mu}_m \cdot \frac{1}{g} \overrightarrow{\mathcal{D}\Phi}^a  \label{eq3_36c} \\
\delta^2 A^a_{5} &=    2\vec{\mu}_e \cdot \frac{1}{g} \overrightarrow{\mathcal{D}\Phi}^a  \label{eq3_36d} \\
\delta \lambda^a_{ -1} &=\sum_{\mathfrak{a}} 2\: \slashed{\mathcal{D}} \Phi\: \Gamma^0\:  [ \cos \psi + \sin \psi\: (-i \Gamma^{(4)})]\:   \zeta_{\mathfrak{a}, -1}\:  .  \label{eq3_36f}
\end{align}
\end{subequations}
Here, the electric dipole moment, due to the spinning magnetic charge, is 
\be
\vec{\mu}_m \equiv \frac{q_m}{2M_{\text{  dyon    }}} \zeta^\dagger_{\mathfrak{a}, -1}  \vec{S}\zeta_{\mathfrak{a}, -1}\: , \label{eq3_38}
\ee
and the magnetic dipole moment, due spinning electric charge, is
\be
\vec{\mu}_e \equiv \frac{q_e}{  2M_{\text{ dyon   }}  }  \zeta^\dagger_{\mathfrak{a}, -1} \vec{S}\zeta_{\mathfrak{a}, -1}\: . \label{eq3_37}
\ee
The spin operator is defined in terms of the Lorentz generators of the rotation group, i.e.
\be
S^l \equiv \epsilon^l_{jk} \left(-\frac{i}{4}\right) [\Gamma^j,\Gamma^k]\: . \label{eq3_39}
\ee

Because the supersymmetric spinors, $\zeta_{\mathfrak{a}, -1}$ are eigenstates of $S^x$ (with eigenvalue 1/2), then
\be
\zeta^\dagger_{\mathfrak{a}, -1} \vec{S} \zeta_{\mathfrak{a}, -1} =\zeta^\dagger_{\mathfrak{a}, -1}  S^x  \zeta_{\mathfrak{a}, -1}\: \hat{e}_x =\frac{1}{2}\: \hat{e}_x  \: . \label{eq3_41}
\ee

The complex constants, $a_{-, \pm 1}  $ in $ \zeta_{-, \pm 1}$, \cref{eqa_13} are arbitrary, and, consequently, different sets of dyon solutions are obtained when quantizing the fermionic modes.  Specifically, choosing both $a_{-,  +1}=0  $ or $a_{-, -1}=0  $ results in spin 0 dyon solutions.   Choosing either $a_{-,  +1}=0  $ or $a_{-, -1}=0  $ yields two sets of spin $1/2$ dyons with $S^x=+1/2$.  Alternatively, interchanging $a^*_{-, \pm 1}  $ with $a_{-, \pm 1} $ yields dyon solutions with $S^x=-1/2$.  Setting both constants  not equal to zero, simultaneously, we obtain spin 1 dyon solutions where $S^x= \pm 1, 0$.  Considering all of these dyon solutions in total, we can evaluate  $\vec{\mu}_m $ and $\vec{\mu}_e$, explicitly, where for the spin 0 dyon, $S^x= 0$, for the two spin 1/2 dyons    $S^x= \pm 1/2$, and for the spin 1 dyon   $S^x= \pm 1, 0$.

The potential functions $\delta^2 A^a_{ 0}$ and  $  \delta^2 \overrightarrow{A}^a$ are amenable to straightforward interpretation.  Given that 
\be
\lim_{r \rightarrow \infty} \frac{1}{g} \overrightarrow{\mathcal{D}  \Phi}^a =\frac{1}{r^2}\: \frac{e_r^a}{n_m}\:  , \label{eq3_40}
\ee
then $\delta^2 A^a_{\mathfrak{a} 0}$ and $\delta^2 A^a_{\mathfrak{a} i}$  in the limit of large $r$ approach the classical electric and magnetic dipole potentials. The factor of 2 preceding each dipole moment is the gyromagnetic (``gyroelectric'') ratio.\footnote{Kastor and Na have, previously, obtained the gyroelectric ratio in their analysis of magnetic monopoles within $N=2$ super Yang-Mills theory.\cite{kastord99}}   

It is apparent that the electric dipole field derived from the potential $\delta^2 A^a_{ 0}$ is equal but opposite to the field derived from the potential $\delta^2 A^a_{ 4}$.  Not as obvious is the fact that the magnetic dipole field derived from the potential $\delta^2 \overrightarrow{A}^a$  is also equal but opposite to that derived from the potential $\delta^2 A^a_{ 5}$.  This relationship follows directly from the fact that $\mathcal{D}^i \mathcal{D}_i \Phi^a=0 $. A similar situation occurs in the Maxwell theory in which the magnetic field derived from the vector, dipole potential is, except for a minus sign,  identical in form to the electric field derived from the scalar, dipole potential.

%% file: input_file_4.tex
The purpose of this section is to generalize the results of \cref{sect2_1} to  superstring theory.   As we show, the solutions obtained in \cref{sect2_1} correspond, in superstring theory, to D3-branes, which are embedded in an   M5-brane compactified on a  type IIB torus\cite{shiberg18}. 

First, the arena for discussing the dyon solutions of \cref{sect2_1} is the M5-brane. The M5 brane is a 5+1 hypersurface propagating in D=1+10 dimensions\cite{simonj12}.  The underlying theory  is based on a single copy of D=11 Majorana fermions which in D=10 superstring theory reduces to two Majoriana-Weyl fermions.  The defining characteristic of these  fermions, $\epsilon$, is that they satisfy a constraint equation, i.e.\@ $\kappa$ symmetry,
\be
\Gamma^{(6)}\: \epsilon = \epsilon\: .     \label{eq4_1a}
\ee
This is precisely the constraint placed on the spinors, \cref{eqa_11}, defining the dyon solutions in \cref{sect2_1}.  Consequently, from the perspective of D=11, the dyon solutions obtained previously  live, in fact, on an M5-brane.

The application of supersymmetry to string theory is fraught with significant, non-trivial technical issues.   First, in the case of superstring theory the bosonic part of the action based on the Lagrangian density~\cref{eqa_1} is replaced by the Dp-brane action which is given by the non-abelian Dirac-Born-Infeld  plus Wess-Zumino-like actions\footnote{Note: the antisymmetric tensor $B_{AB}=0$, where the only non-vanishing R-R potential is $C_{(1)}$, which is a constant background.}
\be 
S = S_{\text{DBI}} + S_{\text{WZ}} \:,   \label{eq4_1}    
\ee
where
\begin{equation}
\begin{split}
 S_{DBI} = -\tau_p \int_{\mathcal{M}_{p+1}}  \text{STr} \{ e^{-\Phi} \sqrt{-\text{det}\:  (g  
                               + 2\pi \alpha^\prime F)} \} \label{eq4_2}
\end{split}   
\end{equation}
and
\begin{equation}
S_{WZ} = \mu_p \int_{\mathcal{M}_{p+1}}     P[C_{(1)}]  \wedge \text{STr}\: \{e^{2 \pi \alpha^\prime F }\}   \:. \label{eq4_3}
\end{equation}
Here $\tau_p$ is the physical tension of the Dp-brane,  $\mu_p$ is its R-R charge, and  $g_{\alpha \beta}=P[G_{MN}]$ is the pull-back of the background metric $G_{MN}$. STr indicates a symmetric trace for terms involving products of the generators of the gauge group (See reference~\cite{olszewskie15} and references therein.).  In \cref{eq4_2}, it is known that after expanding the square root as a power series in $F_{AB}$, computation of the symmetric trace yields ambiguous results in terms of order $F^6$\cite{bergshoeffe01, tseytlina97}.

The fermionic action  based on the Lagrangian density~\cref{eqa_1} is replaced by the fermionic, Dp-bane  action
\begin{equation}
 S_{F} = \frac{\tau_p}{2} \int_{\mathcal{M}_{p+1}}   e^{-\Phi} \sqrt{-\text{det}  (g+  
                                2\pi \alpha^\prime F)}  
														 \bar{\theta} (1-\Gamma_{Dp})[(\tilde{M}^{-1})^{\alpha \beta}\Gamma_\beta \mathcal{D}_\alpha -\Delta ] \theta\: ,\label{eq4_4}  
\end{equation}
where $\Delta$ vanishes since spacetime background is flat for the cases we are considering.
Here,
\be
M_{\alpha \beta}= g_{\alpha \beta} + F_{\alpha \beta}\: , \label{eq4_5}
\ee
and $\Gamma_\alpha=P[\Gamma_M]$.
For type IIB D(2n+1)-branes
\be
\Gamma_{D(2n+1)}=\sum_{q+r=n+1} \frac{A_{2n+1}(q,r)}{B(q,r)}\: , \label{eq4_6}
\ee
and for type IIA D(2n)-branes
\be
\Gamma_{D(2n)}=\sum_{q+r=n+1} \frac{A_{2n}(q,r)}{B(q,r)}\: , \label{eq4_6c}
\ee
where\footnote{The $\Gamma_{D(2n)}$ for the type IIA theory and the $\Gamma_{D(2n+1)}$ for the IIB theory differ by a factor of -1 in references \cite{martuccil05} and \cite{simonj12}.  The reason derives from the fact that $\Gamma^{(10)}$, denoted $\Gamma_{(10)}$ in \cite{martuccil05},  is defined with indices raised, whereas in \cite{simonj12} $\Gamma_{(10)}$   is defined with indices lowered. We adopt the same convention for $\Gamma_{(10)}$, as \cite{martuccil05}.}
\be
\begin{split}
A_{2n+1}(q,r)= & (-1)^{r+1} (i\sigma_2) (\sigma_3)^r \\
             &  \epsilon^{\alpha_1 \cdots \alpha_{2q} \beta_1 \cdots \beta_{2r}}  F_{\alpha_1 \alpha_2} \cdots F_{\alpha_{2q-1} \alpha_{2q}}\Gamma_{\beta_1 \cdots \beta_{2r}}\: ,
\end{split}
 \label{eq4_6a}
\ee
\be
A_{2n}(q,r)=(-1)^{r+1} \Gamma^{(10)} \epsilon^{\alpha_1 \cdots \alpha_{2q} \beta_1 \cdots \beta_{2r}}  F_{\alpha_1 \alpha_2} \cdots F_{\alpha_{2q-1} \alpha_{2q}}\Gamma_{\beta_1 \cdots \beta_{2r}}\: , \label{eq4_6d}
\ee
and
\be
B(q,r)=q!(2r)!2^q \sqrt{-\text{det}  (g+ 2\pi \alpha^\prime F)}\:  . \label{eq4_6b}
\ee
Since our interest is the type I $SO(32)$ our focus will be the type IIB theory to which the type I $SO(32)$ is related.  For the type IIB theory $\theta$ is a 64 component double spinor
\be
\theta = \begin{pmatrix}
\theta_1 \\
\theta_2
\end{pmatrix}\: . \label{eq4_7}
\ee
Each $\theta_i, (i=1,2)$ is a 32 component Majorana-Weyl spinor of positive chirality, i.e. $\Gamma_{(10)}\: \theta_i =+1\: \theta_i$. In \cref{eq4_6a} the pauli matrices act on the spinorial index $i$ in $\theta_i$.
For the abelian gauge theory the fermionic action is invariant under $\kappa$ symmetry which acts on fermions
\be
\delta \bar{\theta}=\bar{\kappa}(1+\Gamma_{D(2n+1)})\: ,  \label{eq4_8}
\ee
The action $S_{F}$, \cref{eq4_4}, corresponding to the fermionic sector of the theory, strictly speaking, only applies to abelian gauge theories.  The extension to non-abelian gauge theories is plagued with problems similar to those occurring in the bosonic action.  Specifically, expansion of the square root in terms of the gauge fields yields products of generators of the algebra whose symmetric trace is known to result in inconsistencies  at order $F^2$\cite{bergshoeffe01}.  At first, we  ignore these problems and assume that the action $S_{F}$ applies to the non-abelian theory, in which case $ \mathcal{D}$ corresponds to the gauge covariant derivative of the applicable non-abelian gauge theory.   

We now show that the BPS solutions given  in \cref{sect2_1} are exact solutions of type I $SO(32)$ superstring theory.  Since the type I $SO(32)$ theory is derived from the type IIB theory, we, initially, focus on the type IIB theory.  In \cite{olszewskie15} we have shown that the BPS solutions presented in \cref{sect2_1} are also solutions of the equations of motion derived from the non-abelian DBI action, \cref{eq4_1}, and are therefore solutions of the type IIB theory with the fermionic degrees of freedom equal to zero.   In general, these bosonic solutions are not supersymmetric.  In \cite{simonj12} Sim\'on has shown that  whether such a set of bosonic solutions preserves supersymmetry is equivalent to determining if there exist  supersymmetry transformations $\epsilon$, 
\be
\epsilon = \begin{pmatrix}
\epsilon_1 \\
\epsilon_2
\end{pmatrix}\: ,  \label{eq4_9}
\ee
that preserve the bosonic nature of these solutions, i.e.\@ $\theta$ remains zero, and furthermore, that the bosonic solutions remain unchanged  to first order.  To satisfy the condition that $\theta=0$, the combined $\kappa$ and supersymmetry transformations must vanish, i.e.
\be
s\theta=\delta_\kappa \theta + \epsilon = 0\: . \label{eq4_10}
\ee
Here, the $\kappa$ symmetry transformation is
\be
\delta_\kappa \theta = (1+\Gamma_{D(2n+1)})\kappa\: ,  \label{eq4_11}
\ee
 Sim\'on has shown this condition is satisfied when
\be
\Gamma_{D(2n+1)} \epsilon = \epsilon\: . \label{eq4_12}
\ee
Sim\'on has solved \cref{eq4_12} for a supersymmetric D3-brane configuration, i.e.\@ $n=1$, with an abelian gauge field residing on the brane.  We now show how the solutions obtained by Sim\'on can be straightforwardly extended to the BPS solutions  with non-abelian gauge fields, given in \cref{sect2_1}. 

For $n=1$, \cref{eq4_6} becomes  
\be
\begin{split}
 \Gamma_{D(3)}= & \text{STr}\: \frac{1}{4!\: \sqrt{-\text{det}  (g+ 2\pi \alpha^\prime F)}}\:  \epsilon^{\alpha_0 \ldots \alpha_3} \\
					    & (\Gamma_{\alpha_0 \ldots \alpha_3}i \sigma_2+6 F^a_{\alpha_0 \alpha_1}T^a \Gamma_{\alpha_2 \alpha_3} \sigma_1 +3 F^a_{\alpha_0 \alpha_1}T^a  F^b_{\alpha_2 \alpha_3}T^b i\sigma_2 ).
\end{split}
 \label{eq4_13}
\ee
Substituting \cref{eq4_13} into \cref{eq4_12} and rearranging terms, we obtain
\be
\begin{split}
  \text{STr}\: \sqrt{-\text{det}  (g+  F)}\: I_2\: \epsilon  = & \text{STr}\: [1+\Gamma^i\Gamma_0 \mathcal{D}_i \Phi^a T^a  (\cos \psi\: \sigma_3 +\sin \psi\: \sigma_1) \\  
	& -\Gamma^i \Gamma_0\: \sigma_3\: E^a_{i}T^a +\Gamma^i \Gamma^j E^a_{i}T^a \mathcal{D}_j\Phi^b T^b \\
	&  (\cos \psi\: \sigma_3 +\sin \psi\: \sigma_1) -\Gamma^i \Gamma_0 B^a_i T^a\: \sigma_1  \\
	&  +B^{ai}T^a \mathcal{D}_i\Phi^b T^b (\cos \psi\: \sigma_3 +\sin \psi\: \sigma_1)  \\
	&  B^{ia} T^a E^b_{i} T^b i \sigma_2  ]\: \epsilon\: ,
\end{split}
\label{eq4_14}
\ee
where $I_2$ is the identity matrix in two dimensions.
In transitioning from \cref{eq4_13} to \cref{eq4_14}, we have imposed the projection constraints
\begin{align}
-\Gamma^{(4)}\: \sigma_2\: \epsilon & = \epsilon  \label{eq4_15a}  \\
\Gamma_0 \Gamma_\Phi \epsilon & = \epsilon \label{eq4_15b}
\end{align}
The matrix $\Gamma_\Phi$ is defined
\be
\Gamma_\Phi \equiv \Gamma_4 \cos \psi + \Gamma_5 \sin \psi\: . \label{eq4_16}
\ee
Substituting  \cref{eq22_3} and \cref{eq2_23b} into the square root term in \cref{eq4_14} , we obtain (See \cref{sect9} for details.)
\be
\begin{split}
\text{Tr} \{\sqrt{[1+ \sin^2 \psi ( Z^2_r (T^r)^2 + Z^2_\theta  (T^\theta)^2  + Z^2_\phi (T^\phi)^2 )  ]^2 } \}\: I_2\: \epsilon 
\end{split}
\label{eq4_17}
\ee
where 
\be
\begin{split}
Z_r &=   \mathcal{D}_r\Phi^r    \\
Z_\theta &=  \frac{\mathcal{D}_\theta\Phi^\theta}{r}   \\
Z_\phi &=  \frac{\mathcal{D}_\phi\Phi^\phi}{r \sin \theta} 
\end{split}
\label{eq4_18}
\ee
Making the same substitutions into the terms to the right of the equal sign in \cref{eq4_14}, we obtain
\be
\begin{split}
\text{Tr} \{[1+ \sin^2 \psi ( Z^2_r (T^r)^2 + Z^2_\theta  (T^\theta)^2  + Z^2_\phi (T^\phi)^2 )]\: \} \sigma^2_1 \epsilon 
\end{split}
\label{eq4_19}
\ee
The reduction of the right-hand side of \cref{eq4_14} proceeds, for the most part, as in \cite{simonj12} without requiring that the symmetric trace condition, with one exception.  The second term in the second line of \cref{eq4_14} requires invoking the symmetric trace condition to vanish.   
In obtaining both \cref{eq4_17} and \cref{eq4_19} we have used the fact 
\be
\begin{split}
B^{ia}T^a= & B_i^aT^a=\sin \psi Z_{i}T^{a} \delta_i^a \\
E^{ia}T^a= & E_i^aT^a=\cos \psi Z_{i}T^{a} \delta_i^a
\end{split}
\label{eq4_20}
\ee
Using $Z_i$, defined in \cref{eq4_18}, rather than  $\mathcal{D}\Phi_i^a \delta^i_a$, in \cref{eq4_17}, is equivalent, geometrically, to transforming from the orthogonal basis vectors of spherical polar coordinates $(\partial_r, \partial_\theta, \partial_\phi)$ to the orthonormal basis vectors,  $(\partial_r, \hat{\theta}, \hat{\phi})$, i.e.\@  $ds^2=dr^2+ r^2 d\theta^2+r^2\: \sin^2 \theta\:  d\phi^2 = dr^2+ \hat{\theta}^2+ \hat{\phi}^2$.
The reason for this replacement is to facilitate a comparison of results, presented here,  with those presented in \cite{simonj12}, where the metric tensor is given in an orthonormal basis.

We, now, solve the constraint equations, \cref{eq4_15a} and \cref{eq4_15b} for $\epsilon$, obtaining\footnote{Because we have two independent supercharges,($ \mathfrak{a}= +, -$), there are two solutions for $\epsilon$.  We have omitted  labeling $\epsilon$ with an additional subscript $\mathfrak{a}$ so that the notation is less cluttered. } 
\be
\epsilon = \begin{pmatrix}
\zeta_{\mathfrak{a}, +1} \\
-i \Gamma^{(4)}\: \zeta_{\mathfrak{a}, +1} 
\end{pmatrix}\: ,  \label{eq4_21}
\ee
Up to a phase, which we take to be zero, we find that $\zeta_{\mathfrak{a}, -1}$ (\cref{eqa_13}) is given by
\be
\zeta_{\mathfrak{a}, -1} = \epsilon_2 =  -i \Gamma^{(4)}\: \epsilon_1 = -i \Gamma^{(4)}\: \zeta_{\mathfrak{a}, +1}\: . \label{eq4_22}
\ee

The relationship between the Type IIB theory, ($N=2, D=10$) supersymmetry, discussed here, and relevant Type I theory $SO(32)$, ($N=1, D=10$) supersymmetry, can be gleaned from \cref{eq3_36f} and \cref{eq4_22}.
\footnote{It is worth noting that there is a supersymmetric version of the non-abelian Dirac-Born-Infeld action~\cref{eq4_1}.  It construction is based on a generalization of the principles used here to extend the supersymmetric results of \cref{sect2_1} to superstring theory.  See, for example, the work of Bergshoeff et al. \cite{bergshoeffe00}.}  We define 
\be
\epsilon_\text{Type I}= \cos \psi\: \epsilon_2 - \sin \psi\: \epsilon_1\: \label{eq4_23}
\ee
so that 
\be
\delta \lambda^a_{\mathfrak{a}, -1} = 2\: \slashed{\mathcal{D}} \Phi\: \Gamma^0\:  \epsilon_\text{Type I}\: . \label{eq4_24}
\ee

In summary,   we have shown that the  dyon solutions obtained in \cref{sect2_1} satisfy the $\kappa$ symmetry constraint, \cref{eq4_12}, and are therefore, also,  solutions of Type IIB  (Type I $SO(32)$) superstring theory.

%% file: input_file_5.tex
In this section we apply T-duality transformations to the superstring solutions derived in \cref{sect10} and study the duality between the supersymmetric string theoretic solutions obtained therein and their gravitational analogue.  
Specifically, we apply the  T-duality transformations to spatial dimensions $x^4$ and $x^5$ of the M5-brane, transforming the  gauge potential functions, $A_4^a T^a$ and $A_5^a T^a$,  into  embedding coordinates, $2 \pi \alpha' A_4^a T^a$ and $2 \pi \alpha'A_5^a T^a$.  In order that such transformation  be interpreted, straightforwardly, the potentials should not depend on the coordinates $x^4$ or $x^5$, and furthermore should also commute.  The metric obtained on the two resulting $D3$-branes is derived by pulling back the metric induced by the embedding coordinates\cite{johnsonc03}, i.e.\@ the metric, $g_{\mu \nu}$, is given by
\be
g_{\mu \nu}=\eta_{\mu \nu} + \sum_{M,N = 4}^5 \eta_{M N} \text{STr} (\mathcal{D}_\mu A^a_{ M}T^a\:  \mathcal{D}_\nu A^b_{ N}T^b)\:  .  \label{eq11_3}  
\ee

 After including the back reaction in  the T-duality transformations, we find that the potential functions $A_4^a T^a$ and $A_5^a T^a$, \cref{eq3_1}, in general, do not commute complicating their interpretation as embedding coordinates.  Since the non-commutativity is present only for solutions with non-vanishing spin, we organize this section into two subsections, the first dealing with the case of vanishing spin and second dealing with the more complicated case of non-vanishing spin, which includes both spin 1/2 and spin 1 solutions.

\subsection{Case 1: Spin 0 Solutions}
\label{sect11_2}
\begin{figure}[ht]
 \centering
 \includegraphics[width=6.0cm]{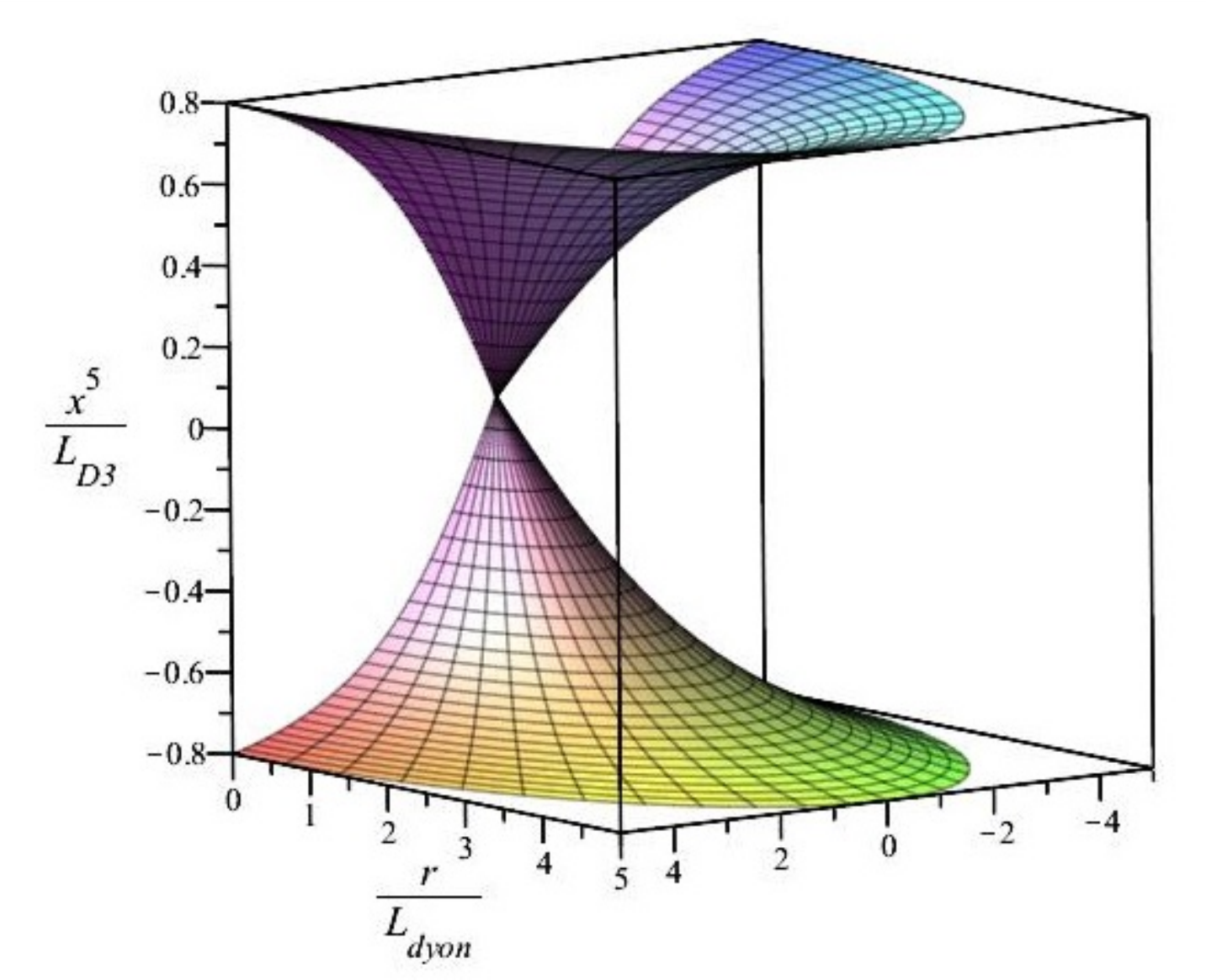}
 \caption{Wormhole.  Shown is the embedding diagram of the two D3-branes with azimuthal angle supressed.  The  radial coordinate,  $r$, has been replaced with the dimensionless coordinate  $ r/L_\text{dyon}$, and the  embedding coordinate $x^5$ has been replaced with the dimensionless coordinate  $   x^5/L_{D3}$. }
\label{fig1}
 \end{figure}

Before applying T-duality transformations, we perform a coordinate transformation in the $x^4, x^5$ plane which induces a gauge transformation on $A^a_{ 4}$ and $A^a_{ 5}$,thereby eliminating $A^a_{ 4}$.   Since, for the spin 0 case (See \cref{eq3_2}.),
\begin{subequations}
\label{eq11_13}
\begin{align}
A^a_{ 4} T^a   & =   \Phi^a T^a \cos \psi\: \label{eq11_13a} \\  
A^a_{ 5} T^a  &  =    \Phi^a T^a \sin  \psi \: \label{eq11_13b}\: .  
\end{align}
\end{subequations}
By rotating  the $x^4, x^5$ coordinate axes through an angle $(\pi/2 - \psi)$, we transform $A^a_{ 4}$ and $A^a_{ 5}$,   
\be
\begin{split}
A^a_{ 4} T^a  \rightarrow  & \sin \psi A^a_{ 4} T^a - \cos \psi A^a_{5} T^a = 0 \\  
A^a_{ 5} T^a  \rightarrow  & \cos \psi A^a_{ 4} T^a + \sin \psi A^a_{5} T^a = \Phi^a T^a =  v\: Q(r)\: T^r\: .
\end{split}
 \label{eq11_1} 
\ee
We note that this transformation leaves unchanged the components of the Minkowski metric $\eta_{M N}, (M, N = 4,5)$.
After diagonalizing the matrix $T^r$, we  apply a T-duality transformation to the $x^5$-coordinate axis.  As a consequence, we obtain  two $D3$-branes embedded in a subspace of the M5-brane, where the embedding coordinates of the two $D3$-branes are $x^5 = \pm L_{D3}\: Q(r)$ ($L_{D3} = 2 \pi \alpha' v = 2 \pi \alpha' M_\text{gluon}/g_{D3} $). In addition, the value of the electric or magnetic charge associated with  one $D3$-brane is opposite in sign of the corresponding charge on the other $D3$-brane.  See \cref{fig1}.
The geometrical interpretation of $L_{D3}$ is straightforward.  It is one half of the separation between the $D3$-branes in the asymptotic region of space, i.e.\@ $r \rightarrow \infty$.\footnote{Alternatively, we can transform $A^a_{ 4} T^a  \rightarrow    A^a_{ 0} T^a - A^a_{4} T^a = 0$, which, also, results in the transformation of   the Minkowski metric,  coincidentally, identical in form to \cref{eq11_8}.  Furthermore, the quantity $L_{D3}$, now, depends only on the magnetic charge, $q_m$, and not  on the electric charge, $q_e$.} This is a consequence of the fact that $\lim_{r \rightarrow \infty} Q(r) =1$.  As noted previously $L_\text{dyon}$ is the characteristic size of the dyon (See \cref{eq3_26}.).

Using \cref{eq11_3,eq11_1} we can calculate the metric tensor 
\be
\begin{aligned}
g_{\mu \nu} dx^\mu dx^\nu  = &  -dtdt +
\frac{4 \pi^{2} \tilde{v}^{2} \alpha^{\prime 2} g_{\mathit{D3}}^{2} \sin^{2}\! \psi  \left(\frac{d}{d u}q\! \left(u\right)\right)^{2}+\alpha^\prime        } {     \sin^{2}\! \psi\:  g_{\mathit{D3}}^{2} \tilde{v}^{2}             } du du   \\
&  + \frac{ ( 4 \pi^{2} \tilde{v}^{2} \alpha^\prime  g_{\mathit{D3}}^{2} q\! \left(u\right)^{2} \tilde{w}^2(u)     \sin^{2}\! \psi  + u^{2})\: \alpha^\prime   }{   \sin^{2}\! \psi\:  g_{\mathit{D3}}^{2} \tilde{v}^{2}   } d\theta d\theta \\
& + \frac{     \left(4 \pi^{2} \tilde{v}^{2} \alpha^\prime  g_{\mathit{D3}}^{2} q\! \left(u\right)^{2} \tilde{w}^2(u)            \sin^{2}\! \psi +u^{2}\right) \alpha^\prime     }{\sin^{2}\! \psi\: g_{\mathit{D3}}^{2} \tilde{v}^{2}}  \sin^2\! \theta\: d\phi d\phi\: ,
\end{aligned}
\label{eq11_101}
\ee
where $\tilde{w}(u)=-1+w(u)$ and $\tilde{v} =v \sqrt{\alpha^\prime}$, and the functions $q(u)$ and $w(u)$ are defined in \cref{eq2_26}.  It is straightforward to calculate the scalar curvature. Details of performing this calculation can be found elsewhere~\cite{olszewskie15}. We omit presenting the scalar curvature here, since it comprises a large number of terms and is not amenable to obvious interpretation.  Nonetheless, We can show that the scalar curvature 
\begin{equation}
\lim_{r \rightarrow \infty }  R \rightarrow
    \begin{cases}
        -\frac{24 \pi^{2} \sin\! \left(\psi \right)^{4} g_{\mathit{D3}}^{2} \tilde{v}^{6} L_{\mathrm{dyon}}^{6}}{\alpha^\prime  r^{6}} & \text{if } \psi \ne 0\\
        0 & \text{if } \psi = 0
    \end{cases}
		\label{eq11_18}
\end{equation}
 so that the geometry of each $D3$-brane is asymptotically flat.  Furthermore, we can also show that the scalar curvature, $R$, is finite everywhere.  In particular,  for small values of $r$ the scalar curvature is given by
\be
\begin{aligned}
\lim_{r \rightarrow 0 } R \rightarrow &  \frac{216 \pi^{2}  \sin^4\! \left(\psi \right) g_{\mathit{D3}}^{2} \tilde{v}^{6}}{\left(4 \pi^{2}  \sin^2\! \left(\psi \right)    \tilde{v}^{4}+9\right)^{2} \alpha^\prime} + \\  
        & \left(\frac{r}{L_\text{dyon}}\right)^2 \frac{8  \tilde{v}^{6} g_{\mathit{D3}}^{2} \pi^{2} \sin\! \left(\psi \right)^{4} \left(100 \pi^{2} \sin\! \left(\psi \right)^{2} \tilde{v}^{4}-1089\right)}{5  \left(4 \pi^{2} \sin\! \left(\psi \right)^{2} \tilde{v}^{4}+9\right)^{3} \alpha^\prime}\: .   
\end{aligned}
\label{eq11_2}
\ee
We are constraining the  Yang-Mills coupling constant on the $D3$-branes such that $0 \le g^2_{D3} \le 4\pi$ .  Solutions when    $g^2_{D3} > 4\pi$ are obtained using  weak/strong duality, i.e.\@ the dual theory is obtained by interchanging electric and magnetic charge and letting  $g^2_{D3} \rightarrow (4\pi)/g^2_{D3}$ .

\begin{figure}[ht]
\begin{subfigure}{.5\textwidth}
  \centering
  \includegraphics[width=.8\linewidth]{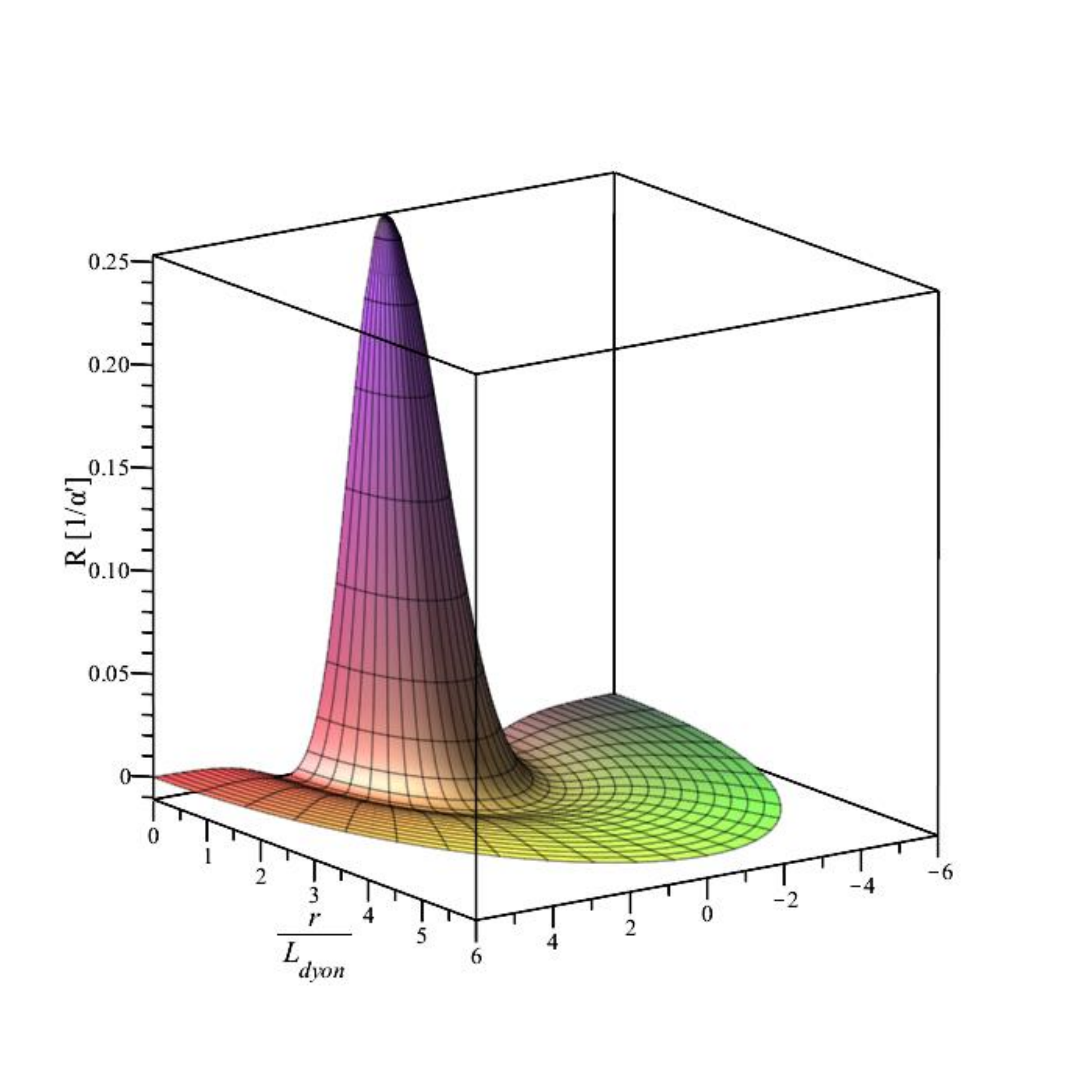}  
  \caption{Magnetic Monopole: The magnetic\\ charge of the monopole is $q_m= 4 \pi$, and\\the electric charge  $q_e=0$, i.e.\@ $n_e=0$ \\ and the $\theta$-term$=0$.  Its mass is $ 4 \pi M_\text{gluon} $.}
  \label{sub1}
\end{subfigure}
\begin{subfigure}{.5\textwidth}
  \centering
  \includegraphics[width=.8\linewidth]{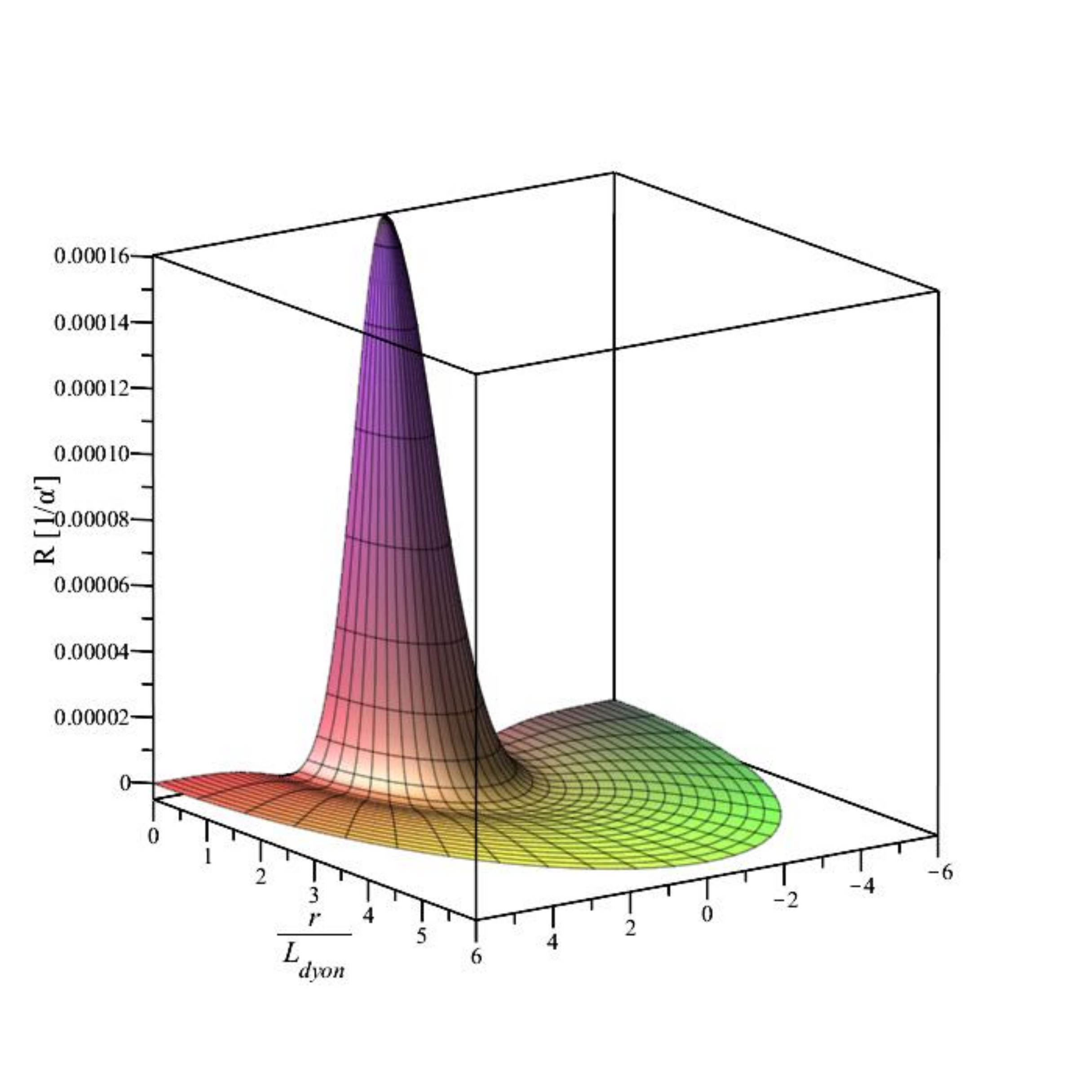}  
  \caption{Dyon: The magnetic charge of the dyon is $q_m= 4 \pi$, and the electric charge  $q_e=1/2$, i.e.\@ $n_e=0$  and the $\theta$-term$=- \pi$.  Its mass is \\ $ \sqrt{1/4 +16 \pi^2} M_\text{gluon} $.}
  \label{sub2}
\end{subfigure}
\caption{Scalar Curvature: Depicted in \cref{sub1,sub2} are the scalar curvatures of a magnetic monopole and dyon, each without spin. In each figure the scalar curvature, $R$, in units of $1/\alpha^\prime$, is plotted as a function of the dimensionless, spherical cooordinates $r/L_\text{dyon}$, and  $\theta$.     The mass of the gluon   $M_\text{gluon} =   (M_P/2)$.  The Yang-Mills coupling constant  $g_{D3}=1$.}
\label{fig2}
\end{figure}
In \cref{fig2} we compare plots of the scalar curvature of a magnetic monopole, with one unit of magnetic charge, to a dyon with one unit of magnetic and one unit of electric charge.  The mass of the gluon $M_\text{gluon} =   M_P/2$ (the Planck mass, $M_P \approx 1/\sqrt{\alpha^\prime}$), and the Yang-Mills coupling constant  $g_{D3}=1$. Note  that the maximum scalar curvature of the dyon is less that that of the monopole.  In addition, it can be shown that when $\psi \rightarrow 0$, i.e.\@  $q_e \rightarrow \infty$, the two $D3$-branes merge into a single $D3$-brane  whose scalar curvature  $R \rightarrow 0$.  Furthermore, it can also be shown, independent of the value of $\psi$, that as $ r \rightarrow \infty$, the metric $g_{\mu \nu} \rightarrow \eta_{\mu \nu}$, and that the geometry of  each $D3$-brane  is asymptotically flat.

\subsection{Case 2: Spin 1/2 and Spin 1 Solutions}
\label{sect11_3}
For the case of non-vanishing spin, the potential functions $A_4^a T^a$ and $A_5^a T^a$ do not commute, except in the spin 1 case when when the x component of the spin, $S^x$, vanishes. We consider, first, the case when $S^x \ne 0$.

For the case of non-vanishing spin
\begin{subequations}
\label{eq11_12}
\begin{align}
A^a_{ 4} T^a   & =   \Phi^a T^a \cos \psi + 2\vec{\mu}_m \cdot \frac{1}{g} \overrightarrow{\mathcal{D}\Phi}^a T^a\: \label{eq11_12a} \\  
A^a_{ 5} T^a  &  =    \Phi^a T^a \sin  \psi + 2\vec{\mu}_e \cdot \frac{1}{g} \overrightarrow{\mathcal{D}\Phi}^a T^a\: \label{eq11_12b}  
\end{align}
\end{subequations}
See \cref{eq3_2,eq3_36c,eq3_36d}.  The non-commutativity of $A^a_{ 4} T^a$ and $A^a_{ 5} T^a $    derives from the fact that $\Phi^a T^a$ and $\hat{e}_x \cdot  \overrightarrow{\mathcal{D}\Phi}^a T^a$ do not commute.  We resolve the problem of non-commutativity by performing the following coordinate transformation,
\be
x^4 \rightarrow x^4 - \cos 2\psi\: x^0 -\sin 2\psi\: x^5\: \label{eq11_9}
\ee
This induces the following gauge transformation
\be
A_4^a T^a \rightarrow A_4^a T^a - \cos 2\psi\: A_0^a T^a -\sin 2\psi\: A_5^a T^a = 0\: , \label{eq11_10}
\ee
thereby eliminating the non-commutivity of $A_4^a T^a$ and $A_5^a T^a$.    
In addition, the metric tensor $\eta_{M N}$ is transformed to
\be
\left( \eta_{M N} \right) \rightarrow \left(\begin{array}{ccc}
-\sin\! \left(2 \psi \right)^{2} & \cos\! \left(2 \psi \right) & \sin\! \left(2 \psi \right) \cos\! \left(2 \psi \right) 
\\
 \cos\! \left(2 \psi \right) & 1 & \sin\! \left(2 \psi \right) 
\\
 \sin\! \left(2 \psi \right) \cos\! \left(2 \psi \right) & \sin\! \left(2 \psi \right) & -\cos\! \left(2 \psi \right)^{2}+2 
\end{array}\right)\: ,
\label{eq11_7}
\ee
for components $M,N = 0, 4, 5$.  The remaining $\eta_{M N}$ are unchanged. 

\begin{figure}[ht]
 \centering
 \includegraphics[width=8.0cm]{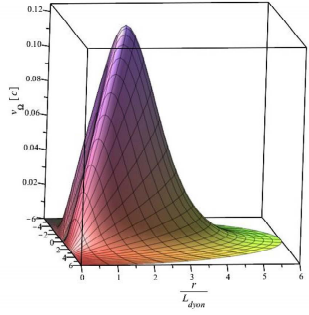}
 \caption{Frame Dragging.  Shown in the plot is the speed,  with which reference frames are dragged around the wormhole relative to a stationary reference frame in a region of space far from the wormhole, i.e.\@ $r \rightarrow \infty$.  The speed, $v_\Omega$, is given as a multiple of the speed of light.  }
\label{fig4}
 \end{figure}

When $S^x \ne 0$ we limit our consideration to solutions where the electric charge, $q_e=0$, i.e.\@ magnetic monopole solutions. Our reason for this limitation is that some calculations, including the  curvature tensor, are calculatingly challenging, and, furthermore, comprise such a large number of terms, that they are not straightforward to interpret. For the magnetic monopole solutions $\psi = \pi/2$ so that  the metric simplifies to
\be
   \left( \eta_{M N} \right) \rightarrow          \left(\begin{array}{ccc}
0 & -1 & 0 
\\
 -1 & 1 & 0 
\\
 0 & 0 & 1 
\end{array}\right)\: .
\label{eq11_8}
\ee
Using \cref{eq11_3,eq11_8,eq11_12b}, we can calculate the metric tensor, $g_{\mu \nu}$.  For these solutions
\be
g_{00} =\frac{A}{B}\: . \label{eq11_102}
\ee
Here
\begin{subequations}
\label{eq11_103}
 \begin{align}
  A  = & 4 \left(-12 \pi^{2} \tilde{v}^{2} \alpha^\prime  g_{\mathit{D3}}^{2} \tilde{w}(u)^{2} q\! \left(u\right)^{2}+u^{2}\right) \times \nonumber \\
    & \times  (S^{x})^2 q\! \left(u\right)^{4} \alpha^{\prime 2} \pi^{2} \tilde{w}(u)^{2} g_{\mathit{D3}}^{6} \sin^{2}\! \theta  \label{eqb11_103a} \\
  B  = & u^{2} \left(4 \pi^{2} \tilde{v}^{2} \alpha^\prime  g_{\mathit{D3}}^{2} \tilde{w}(u)^{2} q\! \left(u\right)^{2}+u^{2}\right)\: . \label{eq11_103b}
 \end{align}
\end{subequations}
The spatial components of the metric, $g_{i j}$, are the same as those of the spin 0 case, \cref{eq11_101} with an important difference in the $\phi$ component of the metric.  Because of the non-vanishing component of the spin $S^x$, reference frames  retrogress about the $x$-axis, so that  the $\phi $ coordinate is transformed
\be
\phi \rightarrow \phi - \Omega\: t\:  .  \label{eq11_104}
\ee
The angular speed, $\Omega$, is
\be
\Omega = -\frac{8 \pi^{2} \alpha^\prime  q\! \left(u\right)^{3}  \tilde{w}^2(u)\{ g_{\mathit{D3}}^{5}\: \tilde{v}^{2} S^x}{ 4 \pi^{2} \alpha^\prime  q\left(u\right)^{2} \tilde{w}^2(u) g_{\mathit{D3}}^{2} \tilde{v}^{2}+u^2}\: . \label{eq11_106}
\ee
Relative to spatial infinity, inertial frames  are dragged with speed
\be
v_\Omega =  L_\text{dyon}\: u \sin \theta\: |\Omega|\: . \label{eq11_107}
\ee
We note that these solutions are, strictly speaking, only accurate to $\mathcal{O} (\alpha^\prime)$, or, equivalently, accurate   for values of the gluon mass, $M_\text{gluon}$, less than the Planck mass, $M_P$. In fact,  we  can show  that whenever  $M_\text{gluon} \lnapprox  M_P$,  the speed with which inertial frames are dragged is less than the speed of light, thus avoiding the possibility of closed time-like curves.    In \cref{fig4} we show a plot of the spatial dependence of $v_\Omega$ for  $M_\text{gluon}=\frac{1}{2} M_P$.

\begin{figure}[ht]
\begin{subfigure}{.5\textwidth}
  \centering
  \includegraphics[width=.8\linewidth]{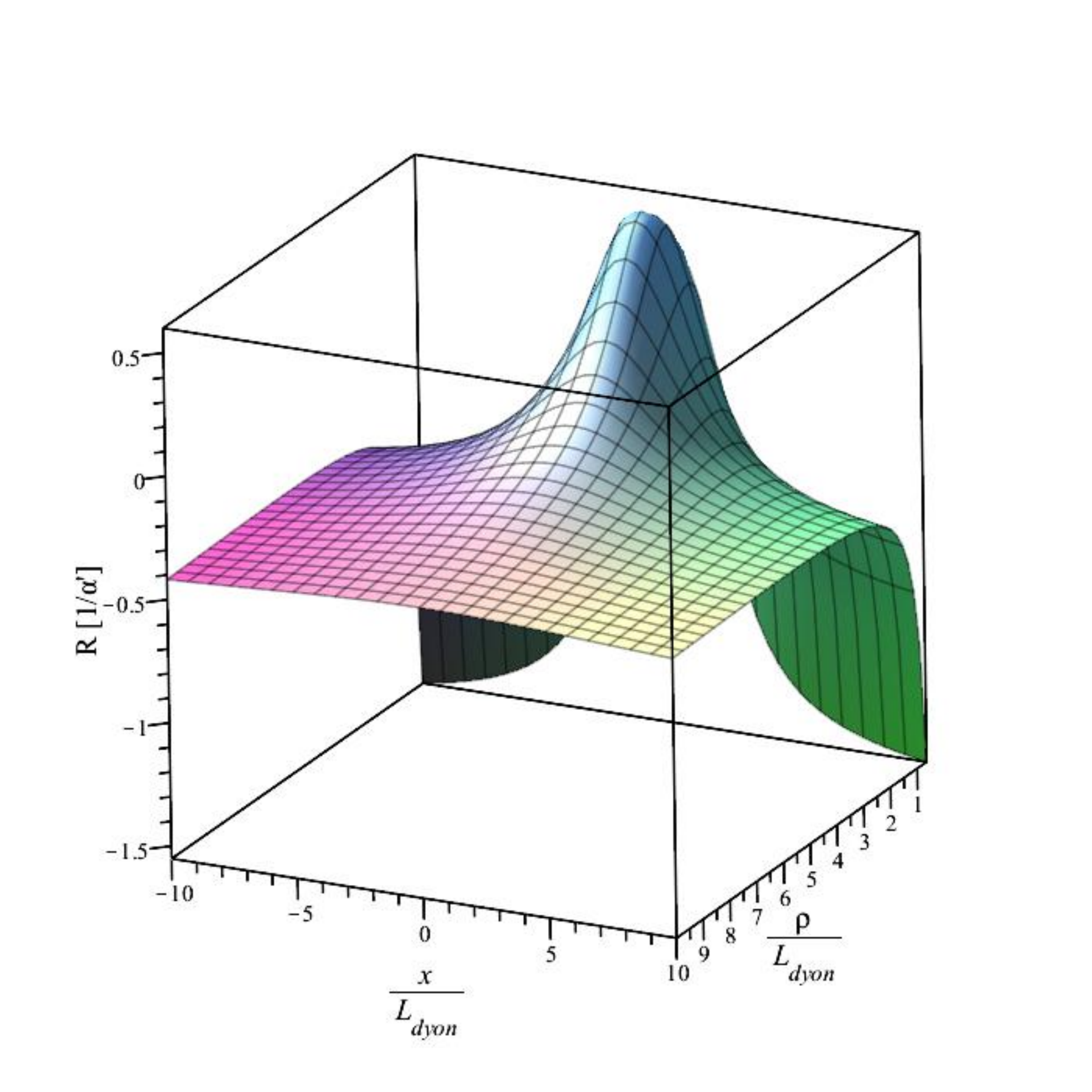}  
  \caption{}
  \label{sub3_1}
\end{subfigure}
\begin{subfigure}{.5\textwidth}
  \centering
  \includegraphics[width=.8\linewidth]{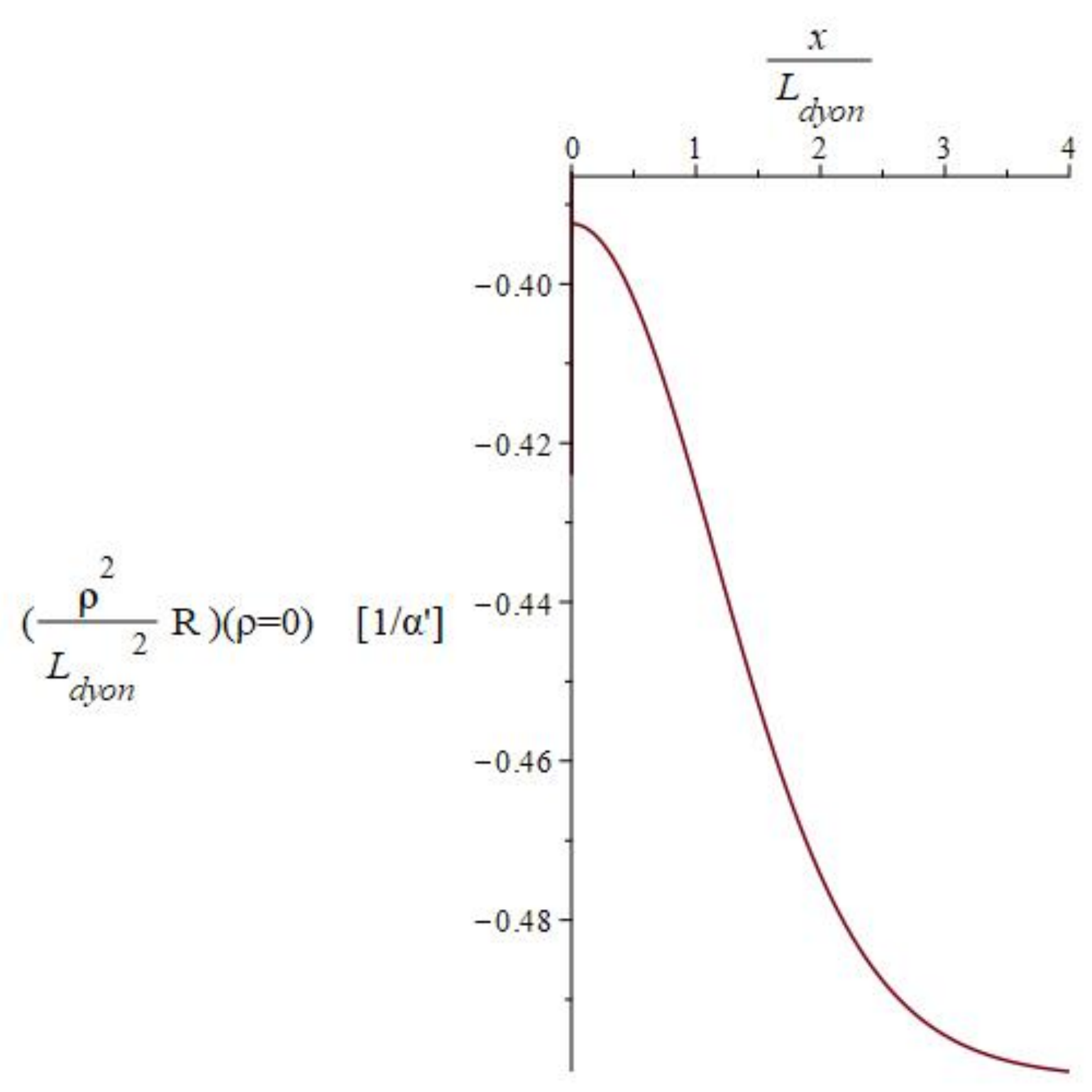}  
  \caption{}
  \label{sub3_2}
\end{subfigure}
\caption{Rotating Magnetic Monopole: In \cref{sub3_1} the scalar curvature, $R$, in units of $1/\alpha^\prime$, is plotted as a function of the dimensionless, cylindrical  cooordinates $\rho/L_\text{dyon}$, and  $x/L_\text{dyon}$.  In \cref{sub3_2} the  scalar curvature, which has been rescaled to remove its divergent behavior, is plotted as a function of $x/L_\text{dyon}$ at $\rho=0$.  The charge of the monopole is $q_m= 4 \pi$, i.e.\@ the electric charge  $q_e=0$ and the $\theta$-term$=0$.       Its mass is    $  4 \pi M_\text{gluon}$.  The Yang-Mills coupling constant  $g_{D3}=1$.}
\label{fig3}
\end{figure}

The scalar curvature in the case of non-vanishing spin is markedly different from that of vanishing spin.  Regarding its general features, we can show that the scalar curvature, $R \propto \frac{1}{\rho^2}$,   as the cylindrical polar coordinate, $\rho \rightarrow 0$.  Here, $\rho \equiv \sqrt{y^2+z^2}$.  In particular, as $r \rightarrow 0$, i.e. $\rho, x \rightarrow 0$, 
\be
\lim_{\rho,x \rightarrow 0} R \rightarrow   -\left(\frac{L_\text{dyon}}{\rho }\right)^2 \frac{18 g_{\mathit{D3}}^{2} \tilde{v}^{2}}{  (  4 \pi^{2}  \tilde{v}^{4}+9 )\alpha ^\prime}\: ,           \label{eq11_5}
\ee
where $\tilde{v} =v \sqrt{\alpha^\prime}$.  Furthermore, we can also show that
\be
\lim_{x \rightarrow \infty} R \rightarrow   -\frac{2 \left(\rho^{2}+L^2_\text{dyon}\right) g_{\mathit{D3}}^{2} \tilde{v}^{2}}{\rho^{2}\: \alpha^\prime}\:     .            \label{eq11_6}
\ee
In \cref{sub3_1} we show a plot of the scalar curvature when  $M_\text{gluon}=\frac{1}{2} M_P$ and $g_{D3}=1$.  In \cref{sub3_2} we show a plot of $\rho^2 R$, evaluated at $\rho=0$.  The purpose of scaling by $\rho^2$ is to remove the divergent part of the scalar curvature.

Surprisingly, the scalar curvature is independent of $S^x$. The reason is that the $R_{00}$ component of the Ricci tensor is the only component which depends on  $S^{ x }$. Specifically, $R_{00} \propto (S^{ x })^2$. In addition, the $g_{00}$ component of the metric tensor is the only component which depends on $S^{ x }$, i.e.\@   $g_{00} \propto (S^{x})^2$.  Thus, after contracting the metric tensor with the Ricci tensor, the  scalar curvature is independent of $S^x$. This is consequential for the spin one monopole solutions when $S^x=0$.  For spin one, when $S^x=0$, \cref{eq11_12a,eq11_12b}    reduce  to \cref{eq11_13a,eq11_13b}, which would seem to indicate that the scalar curvature is the same as for the spin 0 case. Alternatively, as well as preferably, we can  obtain the case  $S^x=0$ as the limit $S^x \rightarrow 0$ for the case of non-vanishing $S^x$.  Taking the limit,  $S^x \rightarrow 0$, we find that $g_{0 0} \rightarrow 0, R_{0 0} \rightarrow 0$, while all other components of the metric tensor, the Ricci tensor, and scalar curvature remain unchanged. This analysis demonstrates that  the case of a spin one, magnetic monopole with $S^x =0$ is inherently different from the that of a magnetic monopole with vanishing spin. 

It is interesting to contrast the geometries of vanishing spin with non-vanishing spin solutions in the asymptotic region of space. 
For the spin zero monopole, as $r \rightarrow \infty$, the metric approaches the Minkowski metric, and the scalar curvature $R \rightarrow 0$.  For the non-vanishing spin monopole, the metric also approaches the Minkowski metric, transformed using ``light cone'' coordinates; however in contrast, as $ \rho \rightarrow \infty$, the scalar curvature
\be
R\rightarrow  -\frac{2  g_{\mathit{D3}}^{2} \tilde{v}^{2}}{ \alpha^\prime}\: . \label{eq11_14}
\ee

%% file: input_file_6.tex
In this study we have investigated the superstring analogue of the 't~Hooft/Polyakov monopole. We have conducted this study in several steps.  First, because superstring theory, naturally, resides in ten dimensions, we have reviewed the dimensional reduction of  $D=10, N=1$ supersymmetry  in a way, that  specifically,  applies to this study. The theory underlying the 't~Hooft/Polyakov monopole is based on a real-valued, scalar boson, which undergoes spontaneous symmetry breaking.  In this study, we assume this boson to be complex-valued so that the monopole (dyon) possesses both magnetic and electric charge. We, next, recast the scalar dyon theory  as  a supersymmetric gauge theory in six dimensions. The complex scalar field is replaced by two real fields which correspond to components of the gauge field in the two extra dimensions.  Applying supersymmetry transformations to the gauge fields, we obtain a theory comprising dyons with spin zero, one half, or one. In addition to possessing both magnetic and electric charge, the dyon possesses both  electric  and magnetic
dipole moments.  We show that both the gyromagnetic and gyroelectric ratio  of each is exactly two, as would be expected \cite{giannakisi97}.\footnote{  The gyroelectric ratio has been reported, previously, by Kastor and Na.} 
Next, we reinterpret the supersymmetric, dyon solutions as solutions in $N=2$, type IIB superstring theory.  
For  $N=2$, type IIB superstring theory the  number of on-shell bosonic and fermionic degrees of freedom is, in general, unequal, the number of fermionic degrees being sixteen and the number of bosonic degrees being eight.  Supersymmetry requires that the degrees of freedom of each be equal.  To prove that the supersymmetric dyon solutions are also supersymmetric  solutions in $N=2$, type IIB string theory, we show that the solutions satisfy  $\kappa$ symmetry constraint equations, which, when satisfied, remove half of the fermionic degrees of freedom.   We then recast the solutions in the type IIB theory as solutions in the type I $SO(32$ theory. We, next, perform a T-duality transformation on the two components of the gauge field in the two extra dimensions.  The T-duality transformation is complicated by the fact that the two gauge fields do not commute, a complication  we resolve  by  eliminating one of the two components of the gauge field by a judiciously  choosen  coordinate/gauge transformation. The transformed solution is a rotating wormhole joining two $D3$-branes.  The electric or magnetic charge of the dyon associated with each $D3$-brane is opposite in sign to that of the other $D3$-brane.
Finally, we analyze the geometry of the $D3$-branes for two cases, one corresponding to  a dyon with vanishing spin, and the other corresponding to a  magnetic monopole with non-vanishing spin.  For the case of vanishing spin, we calculate the metric tensor  and scalar curvature.  We find that the scalar curvature is finite, everywhere.\footnote{We have obtained comparable results in a previous study \cite{olszewskie15}.} In particular, the scalar curvature vanishes, asymptotically far from the throat of the wormhole.  For the case of non-vanishing spin, we, similarly, calculate  the metric tensor and find  that  the spin of the magnetic monopole causes  frame dragging, \cref{eq11_107}.  We, then, calculate the scalar curvature. Unlike the case of vanishing spin, the scalar curvature diverges along the spin quantization axis.  Specifically, it diverges as $1/\rho^2$, $\rho$ being the radial, cylindrical coordinate.  Also, in contrast to the case of vanishing spin, we find that  as $\rho \rightarrow \infty$, the scalar curvature, on the boundary, approaches a constant, negative value, \cref{eq11_14}.

In summary, we note that the wormhole solutions, obtained in this study, provide an example of a gauge, gravity duality.  Furthermore, because they correspond to BPS states and are based on supersymmetry where quantum corrections are expected to be well controlled, we expect   such quantum corrections  not to modify these solutions in a significant way.  Consequently, the underlying theoretical principles may provide some insight in formulating a theory of quantum gravity.

%% file: input_file_7.tex
No underlying data were collected or produced in this study.

%% file: input_file_ap.tex
In this Appendix we present a heuristic derivation of  \cref{eq4_17}.  First, we can show by direct calculation that if  $T^r$, $T^\theta$, and  $T^\phi$  commute then the left side of \cref{eq4_14} evaluates to 
\be
\begin{split}
\text{Tr} \{\sqrt{[1+ \sin^2 \psi ( Z^2_r (T^r)^2 + Z^2_\theta  (T^\theta)^2  + Z^2_\phi (T^\phi)^2 )  ]^2 } \}\: I_2\: \epsilon\:.  
\end{split}
\label{eqc_17}
\ee
On the other hand, if the $T^r$, $T^\theta$, and  $T^\phi$  do not commute there are additional terms.  For example, one such term is proportional to $\sin^6 \psi$,
\be
\begin{split}
& -Z_r^{2} Z_\theta^{2} Z_\phi^{2} ({\textcolor{black}{\mathit{T^\phi}}}^{2} \textcolor{black}{\mathit{T^\theta}} \textcolor{black}{\mathit{T^r}} \textcolor{black}{\mathit{T^r}} \textcolor{black}{\mathit{T^\theta}}-\textcolor{black}{\mathit{T^\theta}} \textcolor{black}{\mathit{T^\phi}} \textcolor{black}{\mathit{T^r}} \textcolor{black}{\mathit{T^\theta}} \textcolor{black}{\mathit{T^\phi}} \textcolor{black}{\mathit{T^r}}  
-\textcolor{black}{\mathit{T^\phi}} \textcolor{black}{\mathit{T^\theta}} \textcolor{black}{\mathit{T^\theta}} \textcolor{black}{\mathit{T^r}} \textcolor{black}{\mathit{T^r}} \textcolor{black}{\mathit{T^\phi}} \\
&+{\textcolor{black}{\mathit{T^\theta}}}^{2} \textcolor{black}{\mathit{T^r}} \textcolor{black}{\mathit{T^\phi}} \textcolor{black}{\mathit{T^\phi}} \textcolor{black}{\mathit{T^r}}+{\textcolor{black}{\mathit{T^r}}}^{2} \textcolor{black}{\mathit{T^\theta}} \textcolor{black}{\mathit{T^\phi}} \textcolor{black}{\mathit{T^\phi}} \textcolor{black}{\mathit{T^\theta}}-{\textcolor{black}{\mathit{T^r}}}^{2} {\textcolor{black}{\mathit{T^\theta}}}^{2} {\textcolor{black}{\mathit{T^\phi}}}^{2}) \sin^{6} \psi
\end{split}
  \:. \label{eqc_1}
\ee
After expanding the square root, we obtain products of such terms. In applying the symmetric trace condition to expressions like \cref{eqc_1}, we first symmetrize each of the terms containing the the $T^r$, $T^\theta$, and  $T^\phi$ with respect to their superscript index.  This effectively makes such terms commute so that the terms vanish.  Consequently, the expression for the square root reduces to \cref{eq4_17}.